\documentclass[prd,twocolumn,showpacs,preprintnumbers,superscriptaddress]{revtex4-1}

\usepackage{mathtools} 
\usepackage{graphicx}
\usepackage{epsfig}
\usepackage{xspace}
\usepackage{dcolumn}
\usepackage{multirow}
\usepackage{longtable}
\usepackage{breqn}
\usepackage{hyperref} 
\usepackage{cleveref}

\def\Journal#1#2#3#4{{#1} {\bf #2}, #3 (#4)}

\def\NIMA{{Nucl.~Instrum.~Methods}~A}

\def\PLB{{Phys.~Lett.}~B}
\def\PRL{Phys.~Rev.~Lett.}
\def\PRD{{Phys.~Rev.}~D}
\def\PRC{{Phys.~Rev.}~C}

\def\PhRep{{Phys.~Rept.}}
\def\APJSUP{{Astrophys.~J.~Suppl.}}
\def\APJ{{Astrophys.~J.}}
\def\JCAP{{J.~Cosmology~Astropart.~Phys.}}
\def\EPJC{{Eur.~Phys.~J.}~C}

\newcommand{\nue}         {$\nu_{e}$\xspace}
\newcommand{\numu}        {$\nu_{\mu}$\xspace}
\newcommand{\nutau}       {$\nu_{\tau}$\xspace}

\newcolumntype{d}[1]{D{.}{\cdot}{#1}}
\newcolumntype{.}{D{.}{.}{-1}}
\newcolumntype{,}{D{,}{,}{2}}

\begin{document}

\title{Indirect Search for Dark Matter from the Galactic Center and Halo with the Super-Kamiokande Detector}
\newcommand{\AFFicrr}{\affiliation{Kamioka Observatory, Institute for Cosmic Ray Research, University of Tokyo, Kamioka, Gifu 506-1205, Japan}}
\newcommand{\AFFkashiwa}{\affiliation{Research Center for Cosmic Neutrinos, Institute for Cosmic Ray Research, University of Tokyo, Kashiwa, Chiba 277-8582, Japan}}
\newcommand{\AFFipmu}{\affiliation{Kavli Institute for the Physics and
Mathematics of the Universe (WPI), The University of Tokyo Institutes for Advanced Study,
University of Tokyo, Kashiwa, Chiba 277-8583, Japan }}
\newcommand{\AFFmad}{\affiliation{Department of Theoretical Physics, University Autonoma Madrid, 28049 Madrid, Spain}}
\newcommand{\AFFubc}{\affiliation{Department of Physics and Astronomy, University of British Columbia, Vancouver, BC, V6T1Z4, Canada}}
\newcommand{\AFFbu}{\affiliation{Department of Physics, Boston University, Boston, MA 02215, USA}}
\newcommand{\AFFuci}{\affiliation{Department of Physics and Astronomy, University of California, Irvine, Irvine, CA 92697-4575, USA }}
\newcommand{\AFFcsu}{\affiliation{Department of Physics, California State University, Dominguez Hills, Carson, CA 90747, USA}}
\newcommand{\AFFcnm}{\affiliation{Institute for Universe and Elementary Particles, Chonnam National University, Gwangju 61186, Korea}}
\newcommand{\AFFduke}{\affiliation{Department of Physics, Duke University, Durham NC 27708, USA}}
\newcommand{\AFFfukuoka}{\affiliation{Junior College, Fukuoka Institute of Technology, Fukuoka, Fukuoka 811-0295, Japan}}
\newcommand{\AFFgifu}{\affiliation{Department of Physics, Gifu University, Gifu, Gifu 501-1193, Japan}}
\newcommand{\AFFgist}{\affiliation{GIST College, Gwangju Institute of Science and Technology, Gwangju 61005, Korea}}
\newcommand{\AFFuh}{\affiliation{Department of Physics and Astronomy, University of Hawaii, Honolulu, HI 96822, USA}}
\newcommand{\AFFicl}{\affiliation{Department of Physics, Imperial College London , London, SW7 2AZ, United Kingdom }}
\newcommand{\AFFkek}{\affiliation{High Energy Accelerator Research Organization (KEK), Tsukuba, Ibaraki 305-0801, Japan }}
\newcommand{\AFFkobe}{\affiliation{Department of Physics, Kobe University, Kobe, Hyogo 657-8501, Japan}}
\newcommand{\AFFkyoto}{\affiliation{Department of Physics, Kyoto University, Kyoto, Kyoto 606-8502, Japan}}
\newcommand{\AFFliv}{\affiliation{Department of Physics, University of Liverpool, Liverpool, L69 7ZE, United Kingdom}}
\newcommand{\AFFmiyagi}{\affiliation{Department of Physics, Miyagi University of Education, Sendai, Miyagi 980-0845, Japan}}
\newcommand{\AFFnagoya}{\affiliation{Institute for Space-Earth Environmental Research, Nagoya University, Nagoya, Aichi 464-8602, Japan}}
\newcommand{\AFFkmi}{\affiliation{Kobayashi-Maskawa Institute for the Origin of Particles and the Universe, Nagoya University, Nagoya, Aichi 464-8602, Japan}}
\newcommand{\AFFpol}{\affiliation{National Centre For Nuclear Research, 02-093 Warsaw, Poland}}
\newcommand{\AFFsuny}{\affiliation{Department of Physics and Astronomy, State University of New York at Stony Brook, NY 11794-3800, USA}}
\newcommand{\AFFokayama}{\affiliation{Department of Physics, Okayama University, Okayama, Okayama 700-8530, Japan }}
\newcommand{\AFFosaka}{\affiliation{Department of Physics, Osaka University, Toyonaka, Osaka 560-0043, Japan}}
\newcommand{\AFFox}{\affiliation{Department of Physics, Oxford University, Oxford, OX1 3PU, United Kingdom}}
\newcommand{\AFFqmul}{\affiliation{School of Physics and Astronomy, Queen Mary University of London, London, E1 4NS, United Kingdom}}
\newcommand{\AFFregina}{\affiliation{Department of Physics, University of Regina, 3737 Wascana Parkway, Regina, SK, S4SOA2, Canada}}
\newcommand{\AFFseoul}{\affiliation{Department of Physics, Seoul National University, Seoul 08826, Korea}}
\newcommand{\AFFsheff}{\affiliation{Department of Physics and Astronomy, University of Sheffield, S3 7RH, Sheffield, United Kingdom}}
\newcommand{\AFFshizuokasc}{\affiliation{Department of Informatics in Social Welfare, Shizuoka University of Welfare, Yaizu, Shizuoka, 425-8611, Japan}}
\newcommand{\AFFstfc}{\affiliation{STFC, Rutherford Appleton Laboratory, Harwell Oxford, and Daresbury Laboratory, Warrington, OX11 0QX, United Kingdom}}
\newcommand{\AFFskk}{\affiliation{Department of Physics, Sungkyunkwan University, Suwon 16419, Korea}}
\newcommand{\AFFtokyo}{\affiliation{The University of Tokyo, Bunkyo, Tokyo 113-0033, Japan }}
\newcommand{\AFFtodai}{\affiliation{Department of Physics, University of Tokyo, Bunkyo, Tokyo 113-0033, Japan }}
\newcommand{\AFFtit}{\affiliation{Department of Physics,Tokyo Institute of Technology, Meguro, Tokyo 152-8551, Japan }}
\newcommand{\AFFtus}{\affiliation{Department of Physics, Faculty of Science and Technology, Tokyo University of Science, Noda, Chiba 278-8510, Japan }}
\newcommand{\AFFtoronto}{\affiliation{Department of Physics, University of Toronto, ON, M5S 1A7, Canada }}
\newcommand{\AFFtriumf}{\affiliation{TRIUMF, 4004 Wesbrook Mall, Vancouver, BC, V6T2A3, Canada }}
\newcommand{\AFFtokai}{\affiliation{Department of Physics, Tokai University, Hiratsuka, Kanagawa 259-1292, Japan}}
\newcommand{\AFFtsinghua}{\affiliation{Department of Engineering Physics, Tsinghua University, Beijing, 100084, China}}
\newcommand{\AFFuw}{\affiliation{Department of Physics, University of Washington, Seattle, WA 98195-1560, USA}}
\newcommand{\AFFynu}{\affiliation{Department of Physics, Yokohama National University, Yokohama, Kanagawa, 240-8501, Japan}}
\newcommand{\AFFllr}{\affiliation{Ecole Polytechnique, IN2P3-CNRS, Laboratoire Leprince-Ringuet, F-91120 Palaiseau, France }}
\newcommand{\AFFbari}{\affiliation{ Dipartimento Interuniversitario di Fisica, INFN Sezione di Bari and Universit\`a e Politecnico di Bari, I-70125, Bari, Italy}}
\newcommand{\AFFnapoli}{\affiliation{Dipartimento di Fisica, INFN Sezione di Napoli and Universit\`a di Napoli, I-80126, Napoli, Italy}}
\newcommand{\AFFroma}{\affiliation{INFN Sezione di Roma and Universit\`a di Roma ``La Sapienza'', I-00185, Roma, Italy}}
\newcommand{\AFFpadova}{\affiliation{Dipartimento di Fisica, INFN Sezione di Padova and Universit\`a di Padova, I-35131, Padova, Italy}}
\newcommand{\AFFkeio}{\affiliation{Department of Physics, Keio University, Yokohama, Kanagawa, 223-8522, Japan}}
\newcommand{\AFFwinnipeg}{\affiliation{Department of Physics, University of Winnipeg, MB R3J 3L8, Canada }}
\newcommand{\AFFkcl}{\affiliation{Department of Physics, King's College London, London, WC2R 2LS, UK }}
\newcommand{\AFFwarwick}{\affiliation{Department of Physics, University of Warwick, Coventry, CV4 7AL, UK }}
\newcommand{\AFFral}{\affiliation{Rutherford Appleton Laboratory, Harwell, Oxford, OX11 0QX, UK }}

\AFFicrr
\AFFkashiwa
\AFFmad
\AFFbu
\AFFubc
\AFFuci
\AFFcsu
\AFFcnm
\AFFduke
\AFFllr
\AFFfukuoka
\AFFgifu
\AFFgist
\AFFuh
\AFFicl
\AFFbari
\AFFnapoli
\AFFpadova
\AFFroma
\AFFkcl
\AFFkeio
\AFFkek
\AFFkobe
\AFFkyoto
\AFFliv
\AFFmiyagi
\AFFnagoya
\AFFkmi
\AFFpol
\AFFsuny
\AFFokayama
\AFFosaka
\AFFox
\AFFral
\AFFregina
\AFFseoul
\AFFsheff
\AFFshizuokasc
\AFFstfc
\AFFskk
\AFFtokai
\AFFtokyo
\AFFtodai
\AFFipmu
\AFFtit
\AFFtus
\AFFtoronto
\AFFtriumf
\AFFtsinghua
\AFFwarwick
\AFFuw
\AFFwinnipeg
\AFFynu

%%%%%%%%%%%%%%%%%%%%%%%%%%%%%%%%%%%%%%%%%%%%%%%%%%%%%%%%%%%%%%%%%%%%
%ICRR
\author{K.~Abe}
\AFFicrr
\AFFipmu
\author{C.~Bronner}
\AFFicrr
\author{Y.~Haga}
\AFFicrr
\author{Y.~Hayato}
\AFFicrr
\AFFipmu
\author{M.~Ikeda}
\author{S.~Imaizumi}
\AFFicrr
\author{H.~Ito}
\AFFicrr 
\author{K.~Iyogi}
\AFFicrr 
\author{J.~Kameda}
\AFFicrr
\AFFipmu
\author{Y.~Kataoka}
\AFFicrr 
% ...

%%%%%%%%%%%%%%%%%%%%%%%%%%%%%%%%%%%%%%%%%%%%%%%%%%%%%%%%%%%%%%%%%%%%%

\collaboration{The Super-Kamiokande Collaboration}
\noaffiliation

\date{\today}

\begin{abstract}
We present a search for an excess of neutrino interactions due to dark matter in the form of  Weakly Interacting Massive Particles (WIMPs) annihilating in the galactic center or halo based on the data set of Super-Kamiokande-I, -II, -III and -IV taken from 1996 to 2016. We model the neutrino flux, energy, and flavor distributions assuming WIMP self-annihilation is dominant to $\nu \overline{\nu}$, $\mu^+\mu^-$, $b\overline{b}$, or $W^+W^-$. The excess is in comparison to atmospheric neutrino interactions which are modeled in detail and fit to data. Limits on the self-annihilation cross section $\langle \sigma_{A} V \rangle$ are derived for WIMP masses in the range 1 GeV to 10 TeV, reaching as low as $9.6 \times10^{-23}$~cm$^3$ s$^{-1}$ for 5 GeV WIMPs in $b\bar b$ mode and $1.2 \times10^{-24}$~cm$^3$ s$^{-1}$ for 1 GeV WIMPs in $\nu \bar \nu$ mode. The obtained sensitivity of the Super-Kamiokande detector to WIMP masses below several tens of GeV is the best among similar indirect searches to date.
\end{abstract}


\maketitle

%%%%%%%%%%%%%%%%%%%%%%%%%%%%%%%%%%%%%%%%%%%%%%%%%%
\section{Introduction}
%%%%%%%%%%%%%%%%%%%%%%%%%%%%%%%%%%%%%%%%%%%%%%%%%%

There is compelling evidence that ordinary baryonic matter composes only $5\%$ of the total energy density of the Universe, which is dominated by dark energy ($68\%$) and dark matter ($27\%$) whose nature is unknown~\cite{dm}. Some well-motivated candidates for particle dark matter (DM) arise within supersymmetric extensions of the Standard Model~\cite{wimps, wimps_kamionkowski}. These particles belong to a collective group referred to as Weakly Interacting Massive Particles (WIMPs). The lightest supersymmetric particle, the neutralino ($\chi$), has been considered one of the most promising WIMP candidates. However, the analysis presented here would be also valid for any other dark matter candidate annihilating into standard model particles.

WIMPs present in the galactic halo may be observed directly via elastic scattering off nuclei in detectors~\cite{direct1, direct2} or indirectly through detection of the products of their annihilations (or decays) into standard model particles, including neutrinos~\cite{indirect1, indirect2, indirect3, beacom}. It is expected that dark matter particles will accumulate in massive celestial objects like stars or planets and be bound by their gravitational potentials~\cite{wikstrom}. Previous searches for WIMP-induced neutrinos from the Sun and Earth cores, based on data collected with the Super-Kamiokande (SK) detector, have shown no excess of dark-matter-induced neutrinos over the atmospheric neutrino background~\cite{sk_koun, sk_tanaka, sk_shantanou}.

In this study, we assume that dark matter particles concentrate in the central regions of galaxies, as predicted by many halo models~\cite{nfw,moore,kravtsov,einasto1,einasto2,burkert}. We search for neutrinos from WIMP annihilation in the center of the Milky Way and from its halo.
This search with neutrinos is complementary to other indirect searches relying on annihilation products like $\gamma$, $e^{+}/e^{-}$ or $\bar p$ ~\cite{indirect_comparison}. The experimental advantage of neutrinos is that they travel unimpeded and undeflected from their origin. However, 
their low interaction cross section puts them at a disadvantage relative to these other annihilation products.  

We constrain thermally averaged self-annihilation cross section $\langle\sigma_{A} V\rangle$ for WIMP pair annihilation for their masses from 1~GeV to 10~TeV. This is the first search for WIMPs from the Milky Way based on data acquired by the Super-Kamiokande detector and extending to dark matter masses below 10~GeV.

The Super-Kamiokande detector is nearly 100\% efficient for detecting neutrino interactions above \mbox{$\sim100$~MeV}. The dominant background is atmospheric neutrino interactions, which have been well studied with our detector. We have made precise measurements of the flux, angular distribution and energy spectra of atmospheric neutrinos~\cite{sk_euan}. We also have a detailed model of systematic uncertainties~\cite{skosc}.

Below we present results of two different approaches to analyzing the same data sample, a full fit employing both data and Monte Carlo templates (combined fit) and a method comparing observed event rates from a region around the galactic center to those in a region located in the opposite part of the sky (on-source off-source analysis). 
The first method fits for a simulated WIMP-induced neutrino contribution to the Super-Kamiokande data together with 
the expected atmospheric neutrino background component. Various WIMP annihilation channels and masses are tested. 
All event categories of detected neutrino interactions used in recent atmospheric analyses~\cite{skosc}, binned in angle and momentum, are used in the fit. The second method provides a check that is entirely data-driven. We compare the number of events in a certain angular region around the galactic center (on-source) and in a region of the same size but shifted by $180^{\circ}$ in right ascension (off-source). As the atmospheric neutrino background equally affects both regions, any excess of events in the on-source data would indicate an additional source of neutrinos from the area around the galactic center.  

%%%%%%%%%%%%%%%%%%%%%%%%%%%%%%%%%%%%%%%%%%%%%%%%%%
\section{Dark Matter Annihilation in the Milky Way}
%%%%%%%%%%%%%%%%%%%%%%%%%%%%%%%%%%%%%%%%%%%%%%%%%%

The energy spectrum of neutrinos from the annihilation of dark matter WIMPs and the branching ratio (BR) for their production is unknown and must be modeled. In the following analysis we consider direct annihilation into pairs of neutrinos, $\chi\chi \rightarrow \nu \bar \nu$, as well as annihilation into pairs of $\mu^{+}\mu^{-}$, $b\bar b$ and $W^{+}W^{-}$. Each annihilation mode is considered separately assuming 100$\%$ BR irrespective of the actual nature of DM annihilation. In the $b\bar b$ and $W^{+}W^{-}$ channels, neutrinos are mainly created in decays of mesons produced during the hadronization of the primary annihilation products. In the $\mu^{+}\mu^{-}$ channel neutrinos are produced directly in the decays of the muons. In the $\nu \bar \nu$ channel, a monoenergetic spectrum and equal fluxes of DM-induced neutrinos of every flavor are assumed, and the energy of the neutrinos equals the mass of the annihilating dark matter particles. 

\begin{figure}[!tb]
	\begin{minipage}{3.4in}
		\includegraphics[width=3.4in,keepaspectratio=true,type=pdf,ext=.pdf,read=.pdf]{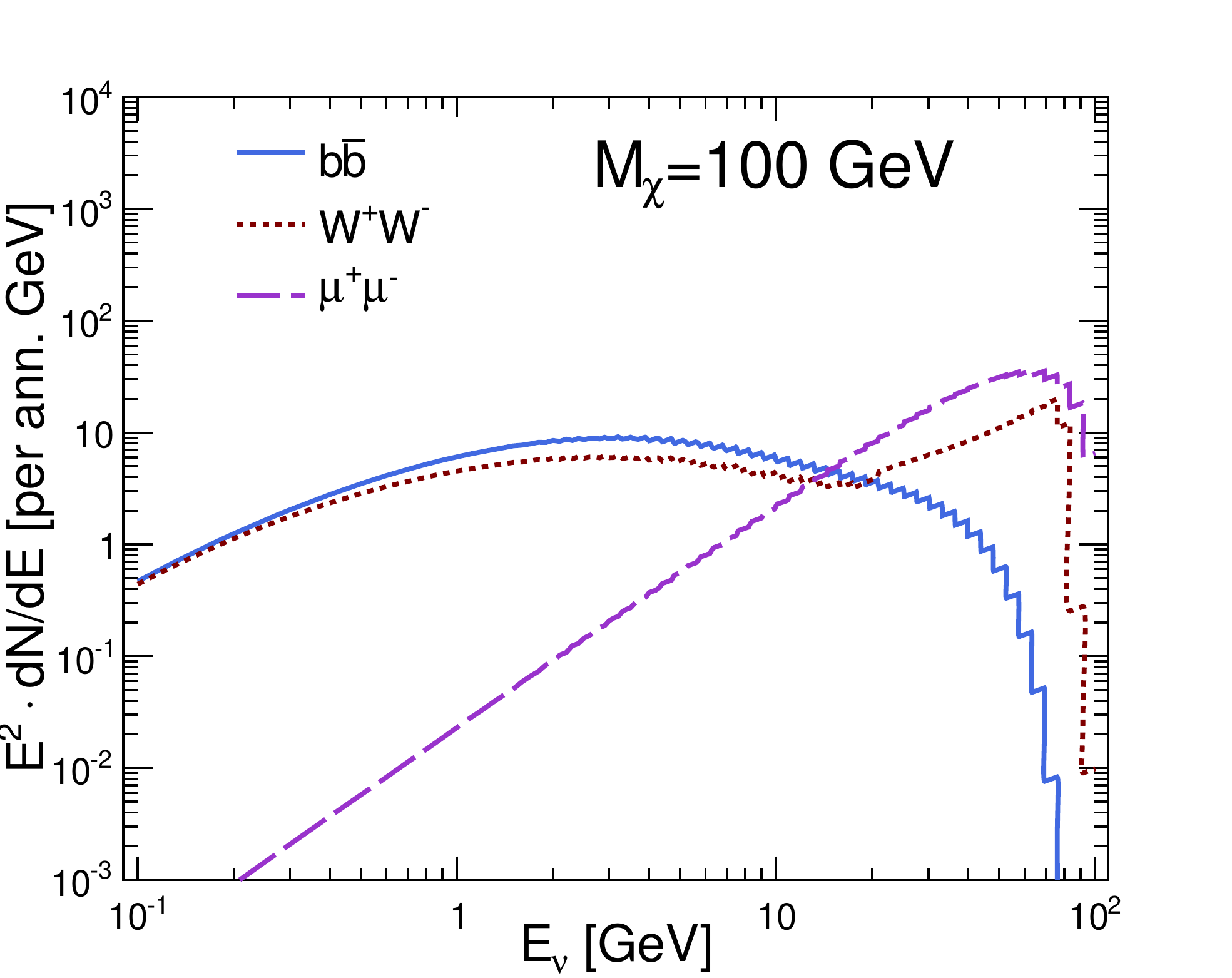}
	\end{minipage}
	\caption{Differential muon neutrino energy spectra for a~WIMP mass of 100~GeV after taking into account neutrino oscillations throughout the galaxy. Fluxes have been calculated based on DarkSUSY~\cite{darksusy} and Eq.~\ref{eq:long_distance_osc_nue},~\ref{eq:long_distance_osc_numu_nutau}.}
	\label{fig:dn_de}
\end{figure}

Neutrinos that travel over galactic distances experience multiple oscillations and
arrive at the detector with a flavor composition that is predictable based on the values of the neutrino mixing parameters~\cite{barenboim,murase}. 
In applying oscillations to DM-induced neutrinos we followed the approach presented in~\cite{icecube22}, adopting parameter values of $\sin^{2}2\theta_{12}=0.86$, $\sin^22\theta_{23}=1.0$ (maximal mixing) and $\theta_{13} \simeq 0 $ to check how long-distance oscillations affect the neutrino production spectra and their flavor ratios at Earth. 
The resulting neutrino fluxes at the detector can be described with the following effective formulas:
%%%%%%%%%%%%%%%%%%%%%%%%%%%%%%%%%%%%%%%%%%%%%%%%%%
\begin{equation}
\phi_{\nu_{e}} \simeq \phi^{0}_{\nu_{e}} - \frac{1}{4} \mathrm{s_{2}}
\label{eq:long_distance_osc_nue}
\end{equation}
%%%%%%%%%%%%%%%%%%%%%%%%%%%%%%%%%%%%%%%%%%%%%%%%%%
and
%%%%%%%%%%%%%%%%%%%%%%%%%%%%%%%%%%%%%%%%%%%%%%%%%%
\begin{equation}
\phi_{\nu_{\tau}} \simeq \phi_{\nu_{\mu}} = \frac{1}{2} ( \phi^{0}_{\nu_{\mu}}+ \phi^{0}_{\nu_{\tau}})+ \frac{1}{8} \mathrm{s_{2}},
\label{eq:long_distance_osc_numu_nutau}
\end{equation}
%%%%%%%%%%%%%%%%%%%%%%%%%%%%%%%%%%%%%%%%%%%%%%%%%%
where $\phi^{0}_{\nu_{e,\mu,\tau}}$ are the initial fluxes, and $\mathrm{s_{2}}$ is defined as $\sin^{2}2\theta_{12}( 2 \phi^{0}_{\nu_{e}} - \phi^{0}_{\nu_{\mu}}-\phi^{0}_{\nu_{\tau}})$~\cite{icecube22}. Note that the above formulas lead to an equal flux of neutrino flavors at the Earth,
that is flux ratios of 1:1:1 for \nue:\numu:\nutau, for the typical scenario of neutrino production via the decays of light mesons and muons in which the initial flavor ratio is 1:2:0 at the production point. The DarkSUSY~\cite{darksusy} simulation package was used to obtain the differential neutrino energy spectra for the DM annihilation channels considered as shown in Fig.~\ref{fig:dn_de} after taking into account neutrino oscillations. 
%%%%%%%%%%%%%%%%%%%%%%%%%%%%%%%%%%%%%%%%%%%%%%%%%%
%%%%%%%%%%%%%%%%%%%%%%%%%%%%%%%%%%%%%%%%%%%%%%%%%%
%%%%%%%%%%%%%%%%%%%%%%%%%%%%%%%%%%%%%%%%%%%%%%%%%%
\begin{figure}[t]
	\begin{minipage}{3.4in}
		\includegraphics[width=3.4in,keepaspectratio=true,type=pdf,ext=.pdf,read=.pdf]{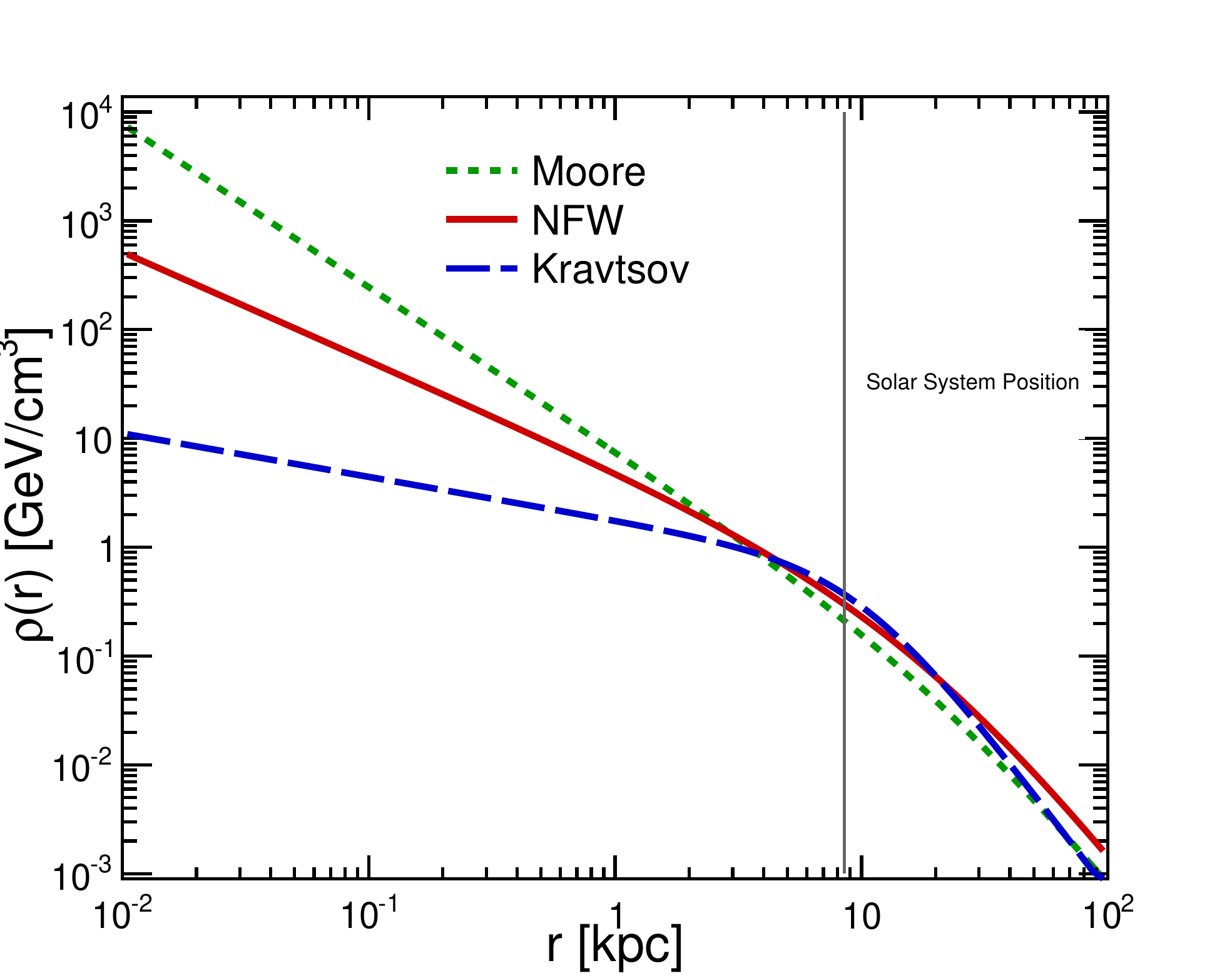}
	\end{minipage}
	\caption{Dark matter density ($\rho$) as a function of distance $(r)$ from the GC  for different DM density distribution profiles: Moore~\cite{moore}, NFW~\cite{nfw} and Kravtsov~\cite{kravtsov}. The vertical gray line indicates the solar system position, $R_{sc} = 8.5$~kpc. The normalizations are chosen to match the local density of DM expected at the position of the solar system from the local rotation curves, for NFW $\rho(R_{sc}) = 0.3{\rm ~GeV~cm^{-3}}$ ($0.27$ for Moore, $0.37$ for Kravtsov)~\cite{beacom}.} %Flat cores are assumed for the inner most regions around the GC.}
	\label{fig:halomodels}
\end{figure}
%%%%%%%%%%%%%%%%%%%%%%%%%%%%%%%%%%%%%%%%%%%%%%%%%%

The expected flux of DM annihilation products depends on the density distribution of DM particles in the Milky Way. 
There are various models describing the structure of the DM halo obtained on the basis of N-body simulations~\cite{nbody} and gravitational lensing observations~\cite{lensing}.  The predictions of the Navarro-Frenk-White (NFW)~\cite{nfw}, the Kravtsov \textit{et~al.}~\cite{kravtsov} and the Moore~\textit{et~al.}~\cite{moore} models of the expected DM density $\rho$ in the halo are shown in Fig.~\ref{fig:halomodels}, assuming a flat DM density profile for the inner most regions of the galaxy. The Moore model anticipates a high density cusp in the center of the Milky Way, while the Kravtsov model yields a flatter density profile. The NFW profile is between these two extreme predictions and is similar to other commonly considered models such as Einasto~\cite{einasto1,einasto2}. In the analysis presented in this paper the NFW model is used as a benchmark.

The DM density distribution in the NFW model can be written as:
%%%%%%%%%%%%%%%%%%%%%%%%%%%%%%%%%%%%%%%%%%%%%%%%%%
\begin{equation}
\rho \left( r \right) = \frac{\rho_{0} }{ \left( r / r_{s} \right)  \left[ 1 + \left( r / r_{s} \right) \right]^{2} },
\label{dmhalo}
\end{equation}
%%%%%%%%%%%%%%%%%%%%%%%%%%%%%%%%%%%%%%%%%%%%%%%%%%
where $r$ is the distance from the center of the galaxy and $r_{s}=20$~kpc is the scale radius. 
The normalization is set so as to match the DM density at the radius of the solar system ($R_{sc} = 8.5$~kpc) expected from rotation curves  and corresponds to $\rho(R_{sc}) = 0.3{\rm ~GeV~cm^{-3}}$ for the NFW profile~\cite{indirect3, beacom}.
%%%%%%%%%%%%%%%%%%%%%%%%%%%%%%%%%%%%%%%%%%%%%%%%%%
\begin{figure}[t]
	\begin{minipage}{3.4in}
		\includegraphics[width=3.4in,keepaspectratio=true,type=pdf,ext=.pdf,read=.pdf]{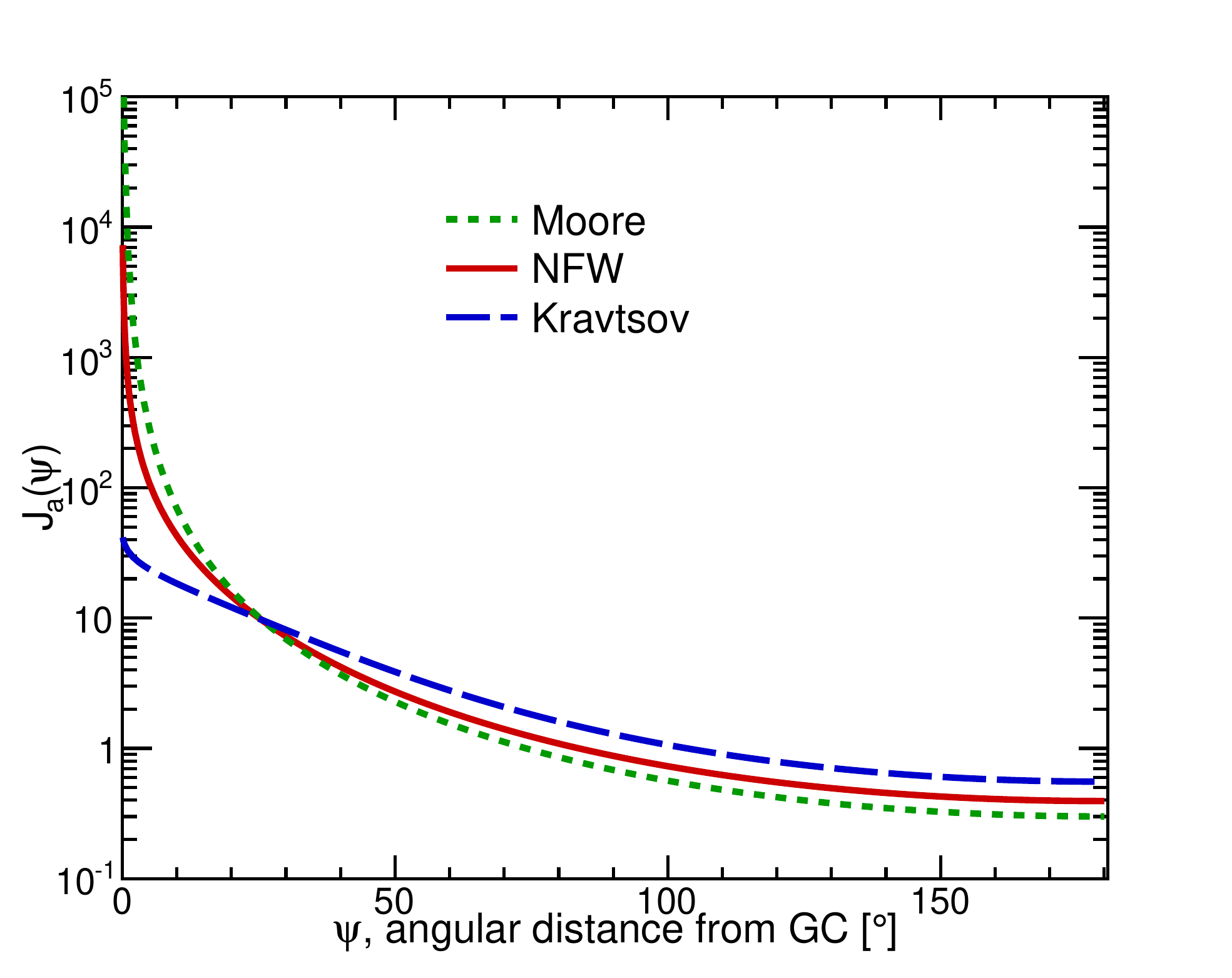}
	\end{minipage}
	\caption{Intensity of annihilation products versus angular distance from the galactic center for
		various DM halo profiles.}
	\label{fig:intensity}
\end{figure}
%%%%%%%%%%%%%%%%%%%%%%%%%%%%%%%%%%%%%%%%%%%%%%%%%%

The annihilation intensity $\mathcal{J}_a$ (numerical flux per solid angle) at an angle $\psi$ with respect to the galactic center (GC) direction is proportional to the line of sight ($l$) integral of the DM density squared (as two DM particles are required for annihilation)~\cite{indirect3, beacom}
\begin{equation}
\mathcal{J}_a(\psi) = \frac{1}{R_{sc} \rho^{2}_{sc}} \int_{0}^{l_{max}} \! \left( \rho \left( \sqrt{R^{2}_{sc} - 2 l R_{sc} \cos{\psi} + l^{2}}\right) \right)^{2} \, \mathrm{d}l,
\label{integral1}
\end{equation}
where $\rho^{2}_{sc} = \rho^{2}(R_{sc})$ and $R_{sc}$ are used as scaling parameters to make $\mathcal{J}_a$ dimensionless. The upper limit of integration,
\begin{equation}
l_{max} =  \sqrt{ ( R^{2}_{MW} -  \sin^{2}{\psi} R^{2}_{sc} ) } + R_{sc} \cos{\psi},
\label{integral2}
\end{equation}
depends on the assumed size of the Milky Way halo, $R_{MW}$. However, contributions beyond the size of the visible stellar halo (20--30 kpc) are negligible. The dependence of $\mathcal{J}_a(\psi)$ on the angular distance from the GC is shown in~Fig.~\ref{fig:intensity}.

The average value of the intensity of DM annihilation products received at the Earth from a cone with half-angle $\psi$ around the GC, that spans a field of view of $\Delta \Omega = 2 \pi (1 - \cos{\psi})$, can be cast as
%In order to estimate the overall intensity of DM annihilation products received from a %given cone half-angle, we define the average value of the intensity in a cone with %half-angle $\psi$ around the GC that spans a field of view of $\Delta \Omega = 2 \pi (1 - \cos{\psi})$ as:
\begin{equation}
\mathcal{J}_a({\Delta \Omega}) = \frac{1}{\Delta \Omega} \int_{ \cos{\psi} }^{1} \! \mathcal{J}(\psi') 2 \pi \, \mathrm{d}(\cos{\psi'}).
\label{integral3}
\end{equation}
Then, the neutrino flux at the Earth can be related to the average intensity of DM annihilation products as~\cite{indirect3, beacom}
\begin{equation}
\frac{\mathrm{d} \phi_{\nu} }{ \mathrm{d} E } =  \frac{\langle\sigma_{A} V\rangle}{2} \mathcal{J}_a ({ \Delta \Omega}) \frac{ R_{sc}\rho^{2}_{sc}}{4 \pi M_{\chi}^{2}} \frac{\mathrm{d} N}{\mathrm{d} E},
\label{dmannihilationrate}
\end{equation}
where $M_{\chi}$ is the assumed mass of the DM particle, the factor $1/2$ is needed for self-conjugate WIMPs, and $1 / 4 \pi$ is for isotropic emission. Here $\mathrm{d} N/ \mathrm{d} E$ is the differential neutrino multiplicity per one annihilation (cf.~Fig.~\ref{fig:dn_de}).

%%%%%%%%%%%%%%%%%%%%%%%%%%%%%%%%%%%%%%%%%%%%%%%%%%
\section{Super-Kamiokande Detector and Data Samples}
\label{sec:datasample}
%%%%%%%%%%%%%%%%%%%%%%%%%%%%%%%%%%%%%%%%%%%%%%%%%%

Super-Kamiokande is a 50~kton water Cherenkov detector located at the Kamioka Observatory operated by the Institute for Cosmic Ray Research of the University of Tokyo~\cite{sk_detector}. It is used to search for proton decay as well as to investigate atmospheric, man-made and extra-terrestrial neutrinos, including solar neutrinos and those produced in supernovae bursts. The detector consists of an inner cylindrical volume (inner detector, ID) and an outer part (outer detector, OD), which plays the role of a veto region for penetrating particles. The detection of neutrinos relies on observation of charged particles, primarily leptons, produced in $\nu$ interactions inside or outside the ID's 22.5~kton fiducial volume. 
Cherenkov radiation from charged particle in the water is emitted in a characteristic conical pattern, is projected onto the detector walls, and is recorded by photomultipliers. The detected light pattern allows for the reconstruction of the particle energy and direction and provides differentiation between electromagnetic showers ($e$-like) and muon-type non-showering particles ($\mu$-like).   

The sample of neutrino interactions in which we search for WIMP-induced neutrinos consists mainly of interactions of atmospheric neutrinos, which are the main background of this search. Based on the topology and energy of the detected events, atmospheric neutrinos  can be assigned to three main categories: fully-contained (FC), partially-contained (PC), and upward-going muons (UP-$\mu$). The FC events have a reconstructed neutrino interaction vertex inside the fiducial volume and particles produced in the parent neutrino interaction stop inside the ID. The true neutrino energy of FC events is in the range of hundreds of MeV up to several GeV. The true neutrino energy of PC events, which have particles exiting into the OD, ranges from 1~GeV to 100~GeV with median energy 10~GeV. Neutrinos of higher energies, 10~GeV to a few TeV, are detected as upward-going muons due to muon neutrino interactions in the surrounding rock. 
Downward-going muons entering from outside the detector are removed from the analysis samples as they cannot be separated from downward-going cosmic ray muons.

%%%%%%%%%%%%% table(data summary) %%%%%%%%%%%%%%%%%%%%%% 
\begin{table*}[htbp]
	\begin{center}
		%\begin{minipage}{9cm}
		\begin{tabular}{l..........}
			\hline \hline
			& \multicolumn{2}{p{2.0cm}}{\rule{0pt}{3ex}}                
			& \multicolumn{2}{p{2.5cm}}{\centering SK-I}                
			& \multicolumn{2}{p{2.5cm}}{\centering SK-II}               
			& \multicolumn{2}{p{2.5cm}}{\centering SK-III}
			& \multicolumn{2}{p{2.5cm}}{\centering SK-IV}
			\\
			& \multicolumn{2}{c}{\rule{0pt}{2ex}}                           
			& \multicolumn{1}{c}{Data}                  
			& \multicolumn{1}{c}{MC}
			& \multicolumn{1}{c}{Data}                  
			& \multicolumn{1}{c}{MC}
			& \multicolumn{1}{c}{Data}                  
			& \multicolumn{1}{c}{MC}
			& \multicolumn{1}{c}{Data}                  
			& \multicolumn{1}{c}{MC} \\
			\hline
			\multicolumn{11}{l}{\rule{0pt}{3ex}Fully Contained (FC) sub-GeV} \\
			\multicolumn{11}{l}{$e$-like, single-ring} \\
			
			\multicolumn{1}{l}{~~~0 decay-e} & \multicolumn{2}{l}{~} &  \multicolumn{1}{r}{2965}  &\multicolumn{1}{l}{2933.9}  &\multicolumn{1}{r}{1577}  &\multicolumn{1}{l}{1541.3}  & \multicolumn{1}{r}{1094}  &\multicolumn{1}{l}{1043.7}  &\multicolumn{1}{r}{4658}  &\multicolumn{1}{l}{4744.9}  \\
			
			\multicolumn{1}{l}{~~~1 decay-e} & \multicolumn{2}{l}{~} &  \multicolumn{1}{r}{299}  &\multicolumn{1}{l}{301.6}  &\multicolumn{1}{r}{168}  &\multicolumn{1}{l}{161.0}  & \multicolumn{1}{r}{117}  &\multicolumn{1}{l}{104.6}  &\multicolumn{1}{r}{590}  &\multicolumn{1}{l}{575.8}  \\

			\multicolumn{9}{l}{$\mu$-like, single-ring} \\        
			
			\multicolumn{1}{l}{~~~0 decay-e} & \multicolumn{2}{l}{~} &  \multicolumn{1}{r}{1017}  &\multicolumn{1}{l}{1008.5}  &\multicolumn{1}{r}{564}  &\multicolumn{1}{l}{550.0}  & \multicolumn{1}{r}{340}  &\multicolumn{1}{l}{356.2}  &\multicolumn{1}{r}{922}  &\multicolumn{1}{l}{907.6}  \\
			
			\multicolumn{1}{l}{~~~1 decay-e} & \multicolumn{2}{l}{~} &  \multicolumn{1}{r}{2006}  &\multicolumn{1}{l}{2046.2}  &\multicolumn{1}{r}{1045}  &\multicolumn{1}{l}{1081.2}  & \multicolumn{1}{r}{744}  &\multicolumn{1}{l}{745.6}  &\multicolumn{1}{r}{4216}  &\multicolumn{1}{l}{4142.4}  \\
			
			\multicolumn{1}{l}{~~~2 decay-e} & \multicolumn{2}{l}{~} &  \multicolumn{1}{r}{148}  &\multicolumn{1}{l}{147.2}  &\multicolumn{1}{r}{86}  &\multicolumn{1}{l}{82.5}  & \multicolumn{1}{r}{58}  &\multicolumn{1}{l}{60.4}  &\multicolumn{1}{r}{395}  &\multicolumn{1}{l}{395.6}  \\
			
			\multicolumn{9}{l}{$\pi^0$-like} \\
			
			\multicolumn{1}{l}{~~~single-ring} & \multicolumn{2}{l}{~}  &  \multicolumn{1}{r}{163}  &\multicolumn{1}{l}{165.4}  &\multicolumn{1}{r}{115}  &\multicolumn{1}{l}{106.0}  & \multicolumn{1}{r}{53}  &\multicolumn{1}{l}{53.4}  &\multicolumn{1}{r}{247}  &\multicolumn{1}{l}{244.7}  \\
			
			\multicolumn{1}{l}{~~~multi-ring} & \multicolumn{2}{l}{~} &  \multicolumn{1}{r}{493}  &\multicolumn{1}{l}{489.0}  &\multicolumn{1}{r}{245}  &\multicolumn{1}{l}{259.0}  & \multicolumn{1}{r}{178}  &\multicolumn{1}{l}{173.6}  &\multicolumn{1}{r}{804}  &\multicolumn{1}{l}{807.0}  \\

			\multicolumn{9}{l}{\rule{0pt}{3ex}Fully Contained (FC) Multi-GeV} \\
			\multicolumn{9}{l}{Single-ring} \\
			
			\multicolumn{1}{l}{~~~$\nu_{e}$-like} & \multicolumn{2}{l}{~} &  \multicolumn{1}{r}{193}  &\multicolumn{1}{l}{168.1}  &\multicolumn{1}{r}{82}  &\multicolumn{1}{l}{84.8}  & \multicolumn{1}{r}{66}  &\multicolumn{1}{l}{60.8}  &\multicolumn{1}{r}{364}  &\multicolumn{1}{l}{359.2}  \\
			
			\multicolumn{1}{l}{~~~$\bar \nu_{e}$-like} & \multicolumn{2}{l}{~} &  \multicolumn{1}{r}{651}  &\multicolumn{1}{l}{663.4}  &\multicolumn{1}{r}{320}  &\multicolumn{1}{l}{339.6}  & \multicolumn{1}{r}{212}  &\multicolumn{1}{l}{232.8}  &\multicolumn{1}{r}{959}  &\multicolumn{1}{l}{949.8}  \\
			
			\multicolumn{1}{l}{~~~$\mu$-like} & \multicolumn{2}{l}{~} &  \multicolumn{1}{r}{699}  &\multicolumn{1}{l}{720.3}  &\multicolumn{1}{r}{400}  &\multicolumn{1}{l}{387.4}  & \multicolumn{1}{r}{237}  &\multicolumn{1}{l}{252.4}  &\multicolumn{1}{r}{1229}  &\multicolumn{1}{l}{1207.2}  \\

			\multicolumn{9}{l}{\rule{0pt}{3ex}Multi-ring} \\
			\multicolumn{1}{l}{~~~$\nu_{e}$-like} & \multicolumn{2}{l}{~} &  \multicolumn{1}{r}{224}  &\multicolumn{1}{l}{228.1}  &\multicolumn{1}{r}{162}  &\multicolumn{1}{l}{155.0}  & \multicolumn{1}{r}{80}  &\multicolumn{1}{l}{85.4}  &\multicolumn{1}{r}{441}  &\multicolumn{1}{l}{442.3}  \\
			
			\multicolumn{1}{l}{~~~$\bar \nu_{e}$-like} & \multicolumn{2}{l}{~} &  \multicolumn{1}{r}{212}  &\multicolumn{1}{l}{215.5}  &\multicolumn{1}{r}{116}  &\multicolumn{1}{l}{124.7}  & \multicolumn{1}{r}{64}  &\multicolumn{1}{l}{67.4}  &\multicolumn{1}{r}{353}  &\multicolumn{1}{l}{366.1}  \\

			\multicolumn{1}{l}{~~~$\mu$-like } & \multicolumn{2}{l}{~} &  \multicolumn{1}{r}{610}  &\multicolumn{1}{l}{615.3}  &\multicolumn{1}{r}{344}  &\multicolumn{1}{l}{336.5}  & \multicolumn{1}{r}{230}  &\multicolumn{1}{l}{226.9}  &\multicolumn{1}{r}{1126}  &\multicolumn{1}{l}{1118.5}  \\
			
			\multicolumn{1}{l}{~~~other } & \multicolumn{2}{l}{~} &  \multicolumn{1}{r}{519}  &\multicolumn{1}{l}{509.4}  &\multicolumn{1}{r}{296}  &\multicolumn{1}{l}{291.7}  & \multicolumn{1}{r}{173}  &\multicolumn{1}{l}{166.5}  &\multicolumn{1}{r}{820}  &\multicolumn{1}{l}{818.4}  \\

			\multicolumn{9}{l}{\rule{0pt}{3ex}Partially Contained (PC)} \\
			
			\multicolumn{1}{l}{~~~Stopping} & \multicolumn{2}{l}{~} &  \multicolumn{1}{r}{141}  &\multicolumn{1}{l}{142.8}  &\multicolumn{1}{r}{81}  &\multicolumn{1}{l}{75.3}  & \multicolumn{1}{r}{55}  &\multicolumn{1}{l}{53.1}  &\multicolumn{1}{r}{289}  &\multicolumn{1}{l}{297.7}  \\
			
			\multicolumn{1}{l}{~~~Through-going}  & \multicolumn{2}{l}{~} &  \multicolumn{1}{r}{793}  &\multicolumn{1}{l}{787.2}  &\multicolumn{1}{r}{356}  &\multicolumn{1}{l}{372.6}  & \multicolumn{1}{r}{308}  &\multicolumn{1}{l}{309.1}  &\multicolumn{1}{r}{1344}  &\multicolumn{1}{l}{1412.1}  \\

			\multicolumn{9}{l}{\rule{0pt}{3ex}Upward-going muons (UP-$\mu$)} \\
			
			\multicolumn{1}{l}{~~~Stopping} & \multicolumn{2}{l}{~} &  \multicolumn{1}{r}{432.3}  &\multicolumn{1}{l}{433.2}  &\multicolumn{1}{r}{203.1}  &\multicolumn{1}{l}{211.1}  & \multicolumn{1}{r}{191.7}  &\multicolumn{1}{l}{169.0}  &\multicolumn{1}{r}{627.1}  &\multicolumn{1}{l}{641.8}  \\
			
			\multicolumn{9}{l}{~~~Through-going} \\
			
			\multicolumn{1}{l}{~~~~~~Non-showering} & \multicolumn{2}{l}{~} &  \multicolumn{1}{r}{1410.6}  &\multicolumn{1}{l}{1349.8}  &\multicolumn{1}{r}{616.4}  &\multicolumn{1}{l}{640.8}  & \multicolumn{1}{r}{508.3}  &\multicolumn{1}{l}{459.5}  &\multicolumn{1}{r}{2497.4}  &\multicolumn{1}{l}{2448.7}  \\
			
			\multicolumn{1}{l}{~~~~~~Showering} & \multicolumn{2}{l}{~} &  \multicolumn{1}{r}{422.0}  &\multicolumn{1}{l}{497.0}  &\multicolumn{1}{r}{193.6}  &\multicolumn{1}{l}{190.0}  & \multicolumn{1}{r}{205.8}  &\multicolumn{1}{l}{239.8}  &\multicolumn{1}{r}{409.2}  &\multicolumn{1}{l}{382.3 \rule[-2ex]{0pt}{0pt}} \\
 
			\hline
			\hline
		\end{tabular}
		\caption{Summary of atmospheric neutrino data and MC event samples in the SK-I, SK-II, SK-III and SK-IV data sets.}
		\label{table:evsummary}
	\end{center}
\end{table*}

%%%%%%%%%%%%%%%%%%%%%%%%%%%%%%%%%%%%%%%%%%%%%%%%%%
\noindent 

\begin{figure*}[thb]
	\begin{minipage}{6.5in}
		\includegraphics[width=6.5in,keepaspectratio=true,type=pdf,ext=.pdf,read=.pdf]{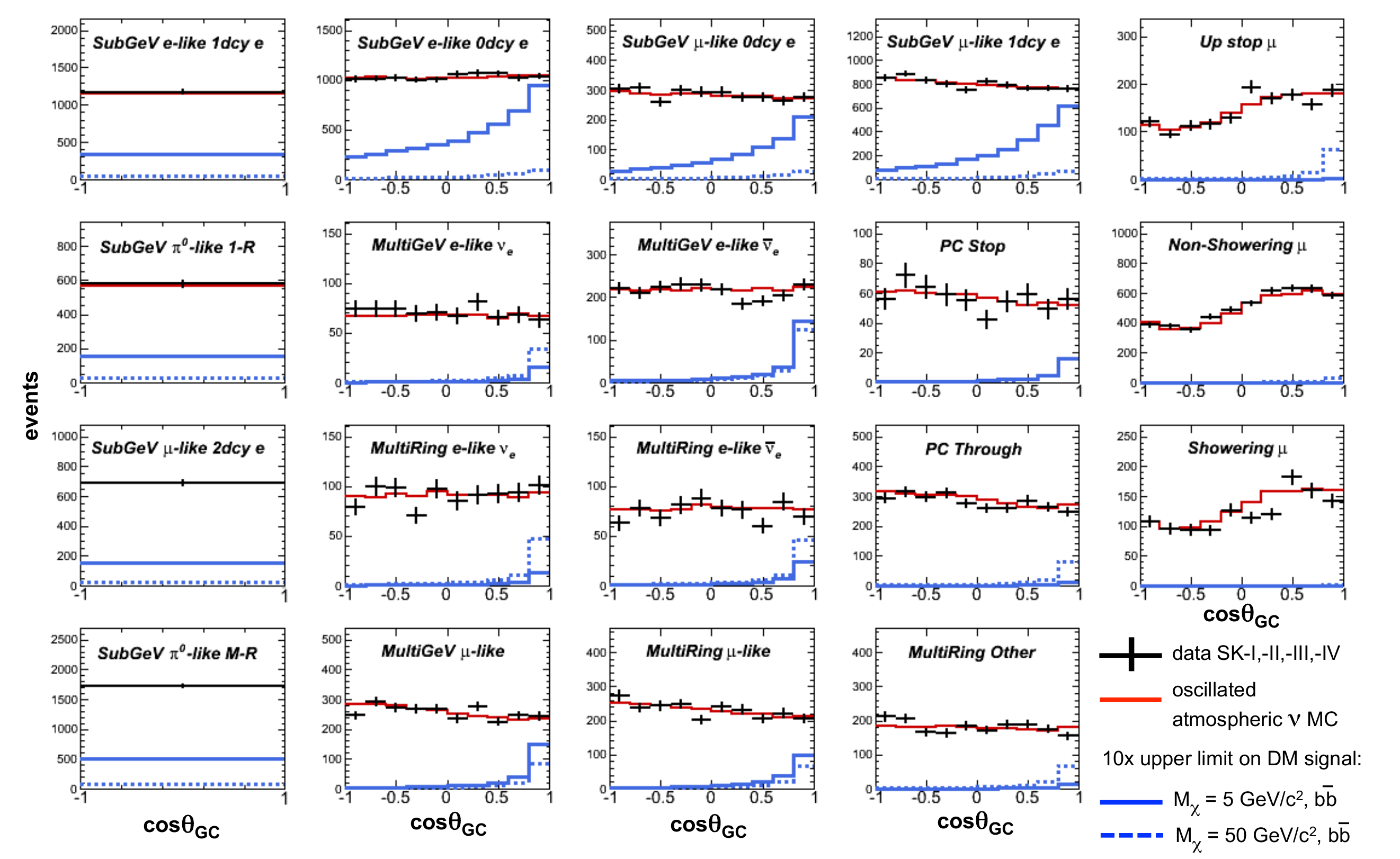}
	\end{minipage}
	\caption{Illustration of the DM annihilation signal for $M_{\chi}=5$~GeV$/$c$^{2}$ (blue solid line) and  $M_{\chi}=50$~GeV$/$c$^{2}$ (blue dotted line) into a pair of $b \bar b$ quarks. The samples used in the fit are presented. SK data (black points with errors),  the ``best-fit'' atmospheric MC after oscillations (red solid) and the DM-induced neutrino signals are shown with respect to the direction of the galactic center ($\cos \theta_{GC}$ = 1 corresponds to the direction of the GC). The signal normalization corresponds to the 90\% C.L. upper limit for the given WIMP mass hypothesis, but has been multiplied by 10 for visibility. In the fitting procedure, the angular distributions shown in the figure are also binned in lepton momentum.}
	\label{fig:sk_ilustr}
\end{figure*}

%%%%%%%%%%%%%%%%%%%%%%%%%%%%%%%%%%%%%%%%%%%%%%%%%%

The FC, PC and UP-$\mu$ event classes can be further divided into more specific subcategories as listed in~Table~\ref{table:evsummary}. The main criteria for dividing FC events are the type of the primary ring ($e$-like vs. $\mu$-like), the number of reconstructed rings (single-ring or multi-ring), the number of electrons from muon decays, the presence of $\pi^{0}$, and the likelihood to be $\bar \nu_{e}$. 
If the reconstructed momentum of the primary lepton is below $1.33$~GeV, the event is classified as sub-GeV, and multi-GeV otherwise. The PC events either stop in or escape the veto region and are thus divided into PC-stopping and PC-through-going. Similarly, UP-$\mu$ events are categorized as ``stopping'' in the detector or ``through-going''. The most energetic muons induce showers while passing through the detector. 
Therefore, ``through-going'' UP-$\mu$ events are further categorized as ``showering'' and ``non-showering''. 
A detailed description of the event classification can be found elsewhere~\cite{skosc}.

Since this search covers a wide range of neutrino energies, all of the SK atmospheric neutrino data and their corresponding subsamples are taken into account in the analysis. The search is based on data from the SK-I (1996-2001), SK-II (2002-2005), SK-III (2006-2008), and SK-IV(2008-2016) run periods. Data used in this work were collected until June 2016 and correspond to a total livetime of 5325.8 live-days for FC and PC events and 5629.1 live-days for UP-$\mu$ events.

The Monte Carlo (MC) simulation of the detector response is based on a GEANT3 model~\cite{GEANT}.
Neutrino interactions are modeled with the NEUT generator~\cite{NEUT} and the background flux is taken from Honda \textit{et al.}~\cite{Honda}.
In this analysis, MC corresponding to 500 years of livetime separately for SK-I, -II, -III and -IV sets (2000 years of livetime in total) is used.

%%%%%%%%%%%%%%%%%%%%%%%%%%%%%%%%%%%%%%%%%%%%%%%%%%
\section{Combined Fit}
\label{sec:analysis}
%%%%%%%%%%%%%%%%%%%%%%%%%%%%%%%%%%%%%%%%%%%%%%%%%%

In this search it is assumed that the neutrino data collected with the SK detector can be described by two components, WIMP-induced neutrinos (the signal) and atmospheric neutrinos (the background). In the combined fit analysis we search for a signal excess in the atmospheric neutrino data sample that is compatible with a neutrino source from the GC by introducing DM signal templates and fitting them together with the atmospheric background assuming a variety of WIMP masses. 
The spectra and directional distributions of signal events are correlated across all of the event subsamples. 
This is different from previous SK searches for DM-induced neutrinos that were based only on the angular information of UP-$\mu$ events~\cite{ sk_tanaka, sk_shantanou}.

Monte Carlo event samples are used to simulate both signal and background. The standard simulation of atmospheric neutrino interactions used in SK's oscillation studies is used to estimate the background and includes simulated tau neutrino interactions~\cite{taupaper}. In order to simulate the signal, a separate and independent sample of events from the atmospheric MC is reweighted to the angular and energy distributions expected for DM-induced neutrinos for a given WIMP mass \mbox{and annihilation} \mbox{channel}.

The Monte Carlo samples allow DM-induced signal simulation for $\nu_\mu \bar \nu_\mu$ with energies $ 100$~MeV$ \le E_{\nu} \le 10$~TeV, for $\nu_e \bar \nu_e$ in $100$~MeV$ \le E_{\nu} \le 80$~GeV and for  $\nu_\tau \bar \nu_\tau$ in $100$~MeV$ \le E_{\nu} \le 150$~GeV. 
These upper limits are determined by event statistics in the atmospheric MC.
For WIMP masses ($M_\chi$) greater than several hundreds of GeV, the detected event rate is dominated by interactions in the rock rather than the detector volume, therefore only muon neutrinos are used to simulate the high energy part of the DM-induced neutrino energy spectrum. Although the analysis in this energy range lacks contributions from other neutrino flavors, it has the advantage of the best pointing resolution.

The signal and background contributions are compared against the data for various WIMP masses and annihilation channels using the ``pull'' $\chi^2$~\cite{pull} method based on Poisson probabilities 
%%%%%%%%%%%%%%%%%%%%%%%%%%%%%%
%% Full ChiSquared def
%
\begin{widetext}
\begin{equation}
\label{eq:fullchi}
\begin{aligned}
&\chi^{2} =  2 \displaystyle \sum_{n} \left( ( N^{ATM}_{n}  ( 1 + \displaystyle \sum_{i} f^{i}_{n} \epsilon_{i} )  +  \beta N^{DM}_{n} ( 1 + \displaystyle \sum_{i} g^{i}_{n} \epsilon_{i} ) ) - N^{data}_{n} + \right.\\ 
& \hspace{3cm} \left. + N^{data}_{n} \ln \frac{ N^{data}_{n} }{ N^{ATM}_{n} ( 1 + \displaystyle \sum_{i} f^{i}_{n} \epsilon_{i} ) + \beta N^{DM}_{n} ( 1 + \displaystyle \sum_{i} g^{i}_{n} \epsilon_{i} )} \right) + \displaystyle \sum_{i} \left( \frac{ \epsilon_{i} }{ \sigma_{i} } \right)^{2},
\end{aligned}
\end{equation}
\end{widetext}

\noindent where $n$ indexes the analysis bins, $N^{ATM}_{n}$ is the atmospheric neutrino background expectation including oscillations,  $N^{DM}_{n}$ is the simulated DM-induced neutrino contribution and $N^{data}_{n}$ is the number of observed events in the $n^{th}$ bin. The parameter $\beta$ represents the normalization of the simulated signal. Determining the ``best-fit'' $\beta$ is the main goal of this analysis. Systematic errors are incorporated into the fit via parameters $\epsilon_{i}$, where $i$ is the systematic error index. 
Here $f^{i}_{n}$ ( $g^{i}_{n}$ ) represents the fractional change in the background (signal) MC in bin $n$ for a $\sigma_{i}$ change in the $i^{th}$ systematic error.
Note that systematic uncertainties related to the detector, data selection and neutrino interactions are fully correlated 
between the signal and background. That is, $f^{i}_{n} = g^{i}_{n}$ for those errors. 
Atmospheric neutrino flux errors, on the other hand, have no impact on the signal and $g^{i}_{n}=0$ for those errors, accordingly.
This analysis only considers DM-induced neutrinos from the GC and halo and does not include other sources. 
During the fit, Eq.~\ref{eq:fullchi} is minimized with respect to the $\epsilon_{i}$ at each point in parameter space according to $ \frac{\partial \chi^{2}}{\partial \epsilon_{i}}= 0$~\cite{pull}. The ``best-fit point'' is defined as the global minimum $\chi^{2}$ on the grid of all tested points. The index $i$ covers systematic uncertainties from ref.~\cite{skosc} considered in the atmospheric neutrino analysis of SK data, but adjusted 
for the GC angular binning scheme used in this analysis.

%%%%%%%%%%%%%%%%%%%%%%%%%%%%%%%%%%%%%%%%%%%%%%%%%%

\begin{figure*}[tb]
	\begin{minipage}{6.5in}
		\includegraphics[width=4.35in,keepaspectratio=true,type=pdf,ext=.pdf,read=.pdf]{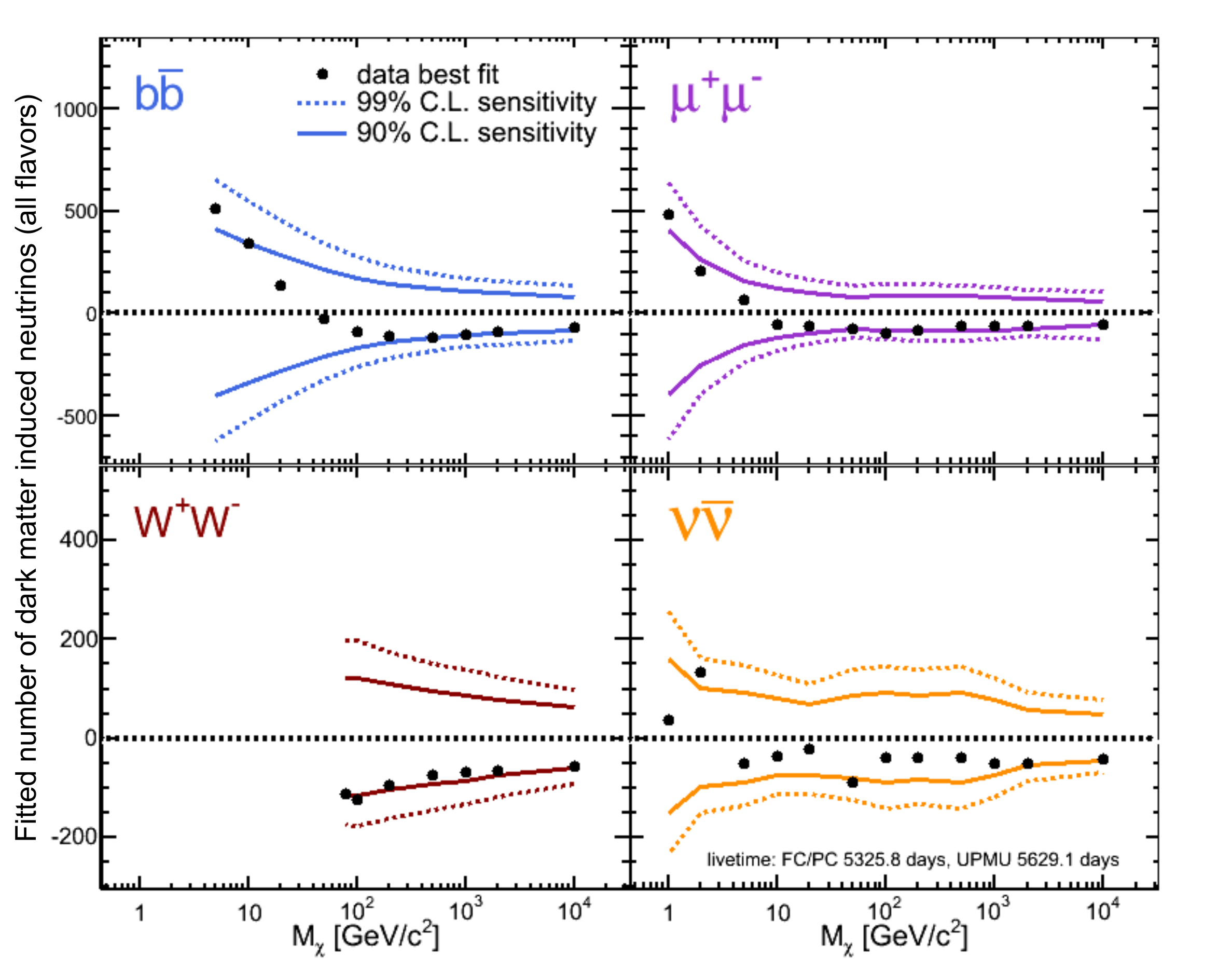}
	\end{minipage}
	\caption{Fitted number of DM-induced neutrinos of all flavors from annihilation into  $b\bar b$, $\mu^{+} \mu^{-}$, $W^{+}W^{-}$ and $\nu\bar \nu$  as a function of the DM mass. Also shown are the expected sensitivities for the zero signal case.
	}
	\label{fig:allowed}
\end{figure*}

%%%%%%%%%%%%%%%%%%%%%%%%%%%%%%%%%%%%%%%%%%%%%%%%%%

The atmospheric neutrino background, $N^{ATM}_{n}$, depends on the values of neutrino oscillation parameters,
  $\theta_{23}$, $\theta_{12}$, $\theta_{13}$, $\Delta m^2_{23}$, $\Delta m^2_{12}$, and $\delta_{\rm CP}$.
The values of these parameters are not determined in this analysis as a given GC coordinate does not correspond to a fixed travel path length in the Earth and therefore provides little sensitivity to neutrino oscillations. 
Accordingly, the oscillation parameters are set to $\delta_{\rm CP}=0$, $\sin^22\theta_{23}=1.0$, $\Delta m^2_{23}=2.44 \times 10^{-3}$ eV$^2$~\cite{t2k}, $\sin^22\theta_{13}=0.092$~\cite{dayabay}, $\sin^2 \theta_{12} = 0.32$, and $\Delta m^2_{12} = 7.46\times10^{-5}$~eV$^{2}$~\cite{solar}. 
However, the uncertainty on each oscillation parameter is included in the fit via systematic error parameters which effectively 
change the $N^{ATM}_{n}$ prediction.

There are 19 data samples used in this analysis, including both $e$-like and $\mu$-like event categories. All the samples listed in Table~\ref{table:evsummary} are used in the search. In total, there are 520 analysis bins ($n=520$) as each sample is binned in momentum and the cosine of the angle between the event direction and the direction of the galactic center ($\cos \theta_{GC}$). The momentum binning is the same as in other SK atmospheric $\nu$ analyses~\cite{skosc}. 
The definition of the event direction depends on the number of rings in the event. 
In the case of the single ring events, the angle to the GC is calculated using the observed Cherenkov ring direction. 
For events with multiple rings, the direction of the event is obtained as the momentum-weighted average direction taken over all ring directions. 
The angular resolution of sub-GeV events strongly depends on the parent neutrino energy and is roughly tens of degrees on average (Fig.~32 in~\cite{skcombinepaper}). At higher energies, the true neutrino direction is more accurately determined due to the high Lorentz boost of the interaction products.

%%%%%%%%%%%%%%%%%%%%%%%%%%%%%%%%%%%%%%%%%%%%%%%%%%
\begin{figure*}[thb]
	\begin{minipage}{6.5in}
		\includegraphics[width=4.25in,keepaspectratio=true,type=pdf,ext=.pdf,read=.pdf]{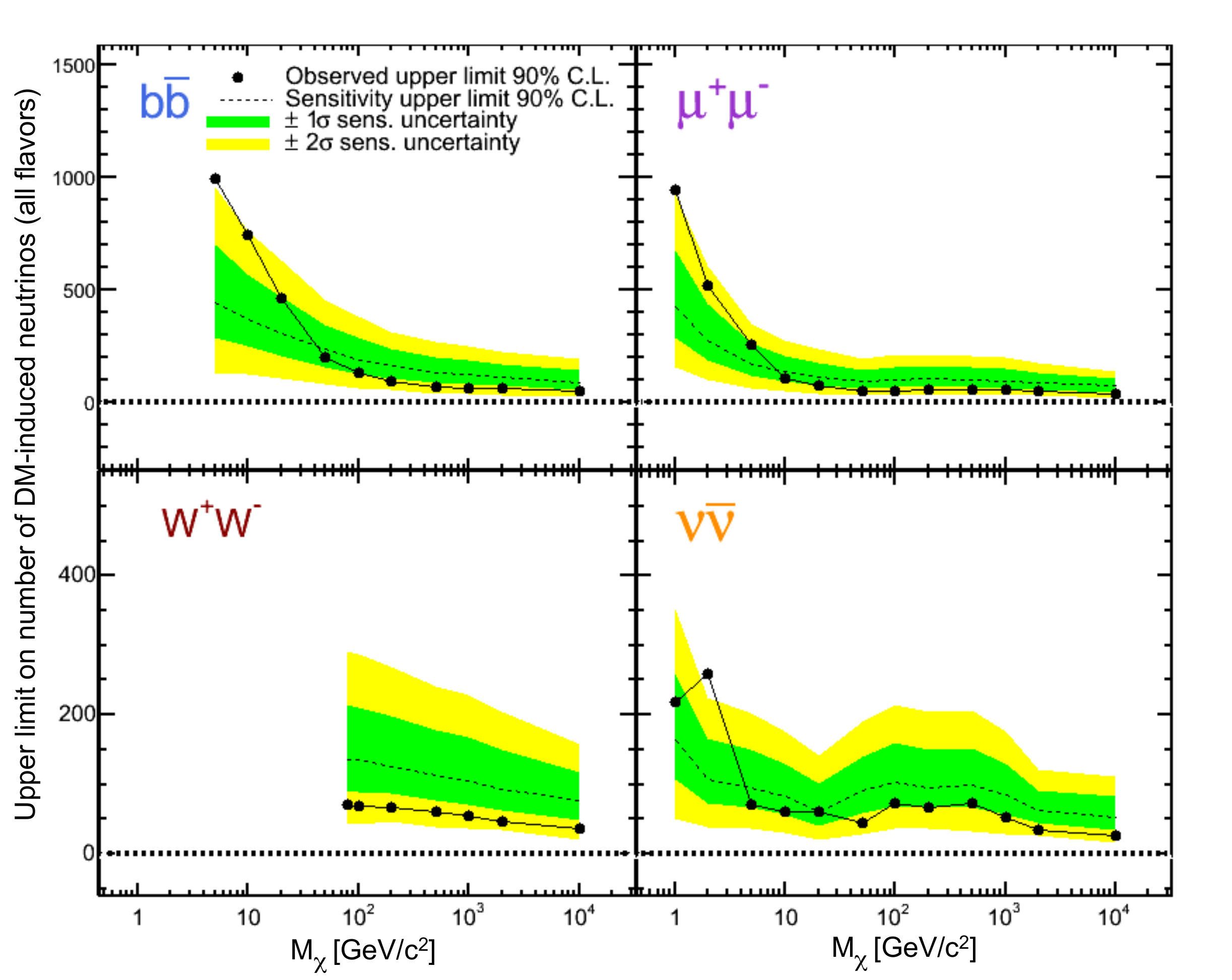}
	\end{minipage}
	\caption{Upper limit on the fitted number of DM-induced neutrinos of all flavors from annihilation into  $b\bar b$, $\mu^{+} \mu^{-}$, $W^{+}W^{-}$ and $\nu\bar \nu$  as a function of the mass of the DM particles. The expected (median) limit assuming no signal is shown by the dashed line and the region containing $68.3\%$ $(95.5\%)$ of the expected limits is shown by the green (yellow) band. %The sensitivity expectation for a null WIMP hypotheis (dashed black line) and their $1\sigma/2\sigma$ uncertainty is presented.
	}
	\label{fig:allowed_brazil}
\end{figure*}
%%%%%%%%%%%%%%%%%%%%%%%%%%%%%%%%%%%%%%%%%%%%%%%%%%

\begin{figure*}[htb]
	\begin{minipage}{6.5in}
		\includegraphics[width=4.31in,keepaspectratio=true,type=pdf,ext=.pdf,read=.pdf]{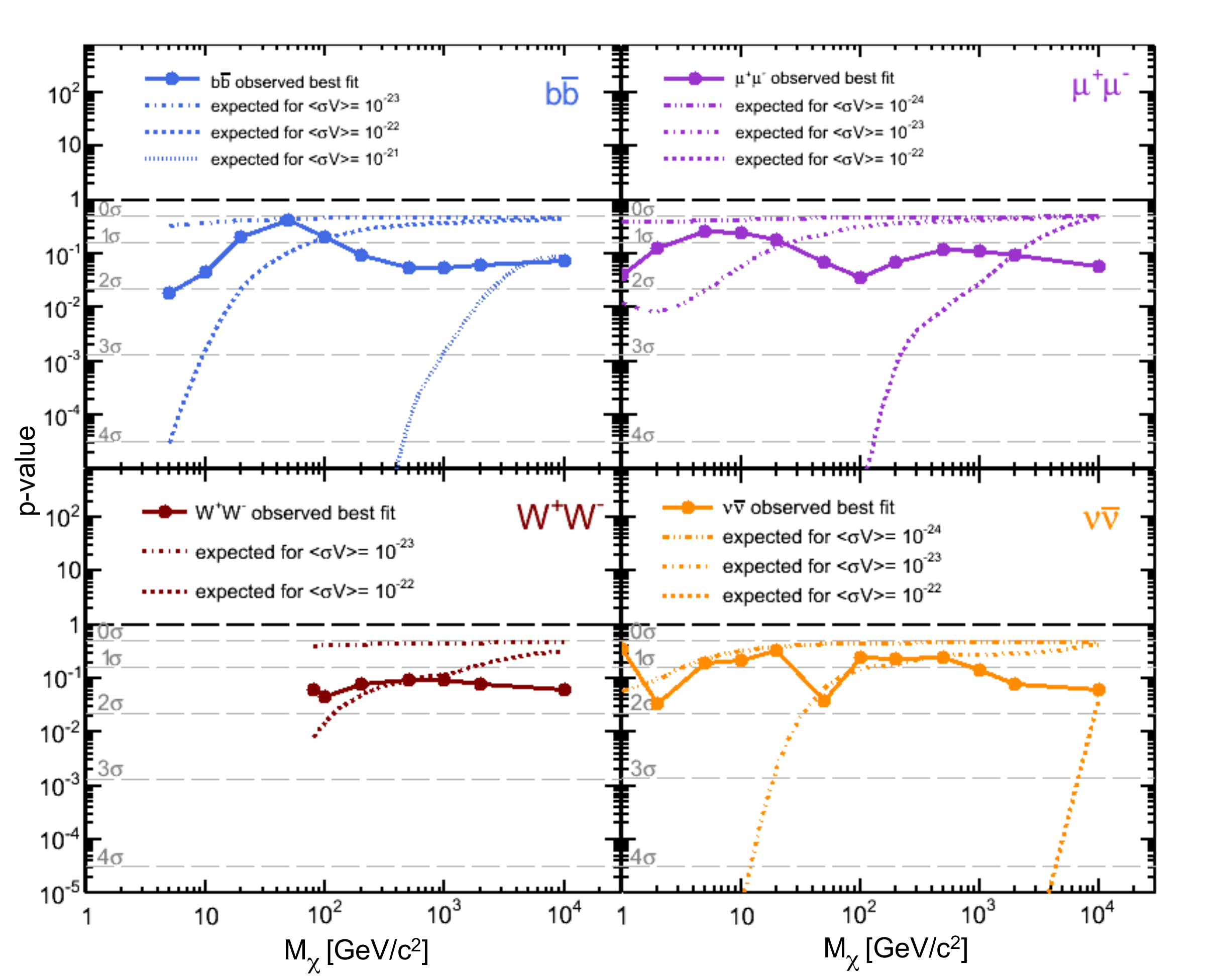}
	\end{minipage}
	\caption{Local $p$-value for the fitted number of WIMP-induced neutrinos for $b\bar b$, $\mu^{+} \mu^{-}$, $W^{+}W^{-}$ and $\nu\bar \nu$ annihilation channels as a function of the mass of the DM particles. Results based on data are shown as solid thick lines with points. For comparison, the expected values assuming different $\langle\sigma_{A} V\rangle$ are also shown.
	}
	\label{fig:pvalue}
\end{figure*}

%%%%%%%%%%%%%%%%%%%%%%%%%%%%%%%%%%%%%%%%%%%%%%%%%%

\begin{figure*}[htb]
	\begin{minipage}{6.5in}
		\includegraphics[width=4.8in,keepaspectratio=true,type=pdf,ext=.pdf,read=.pdf]{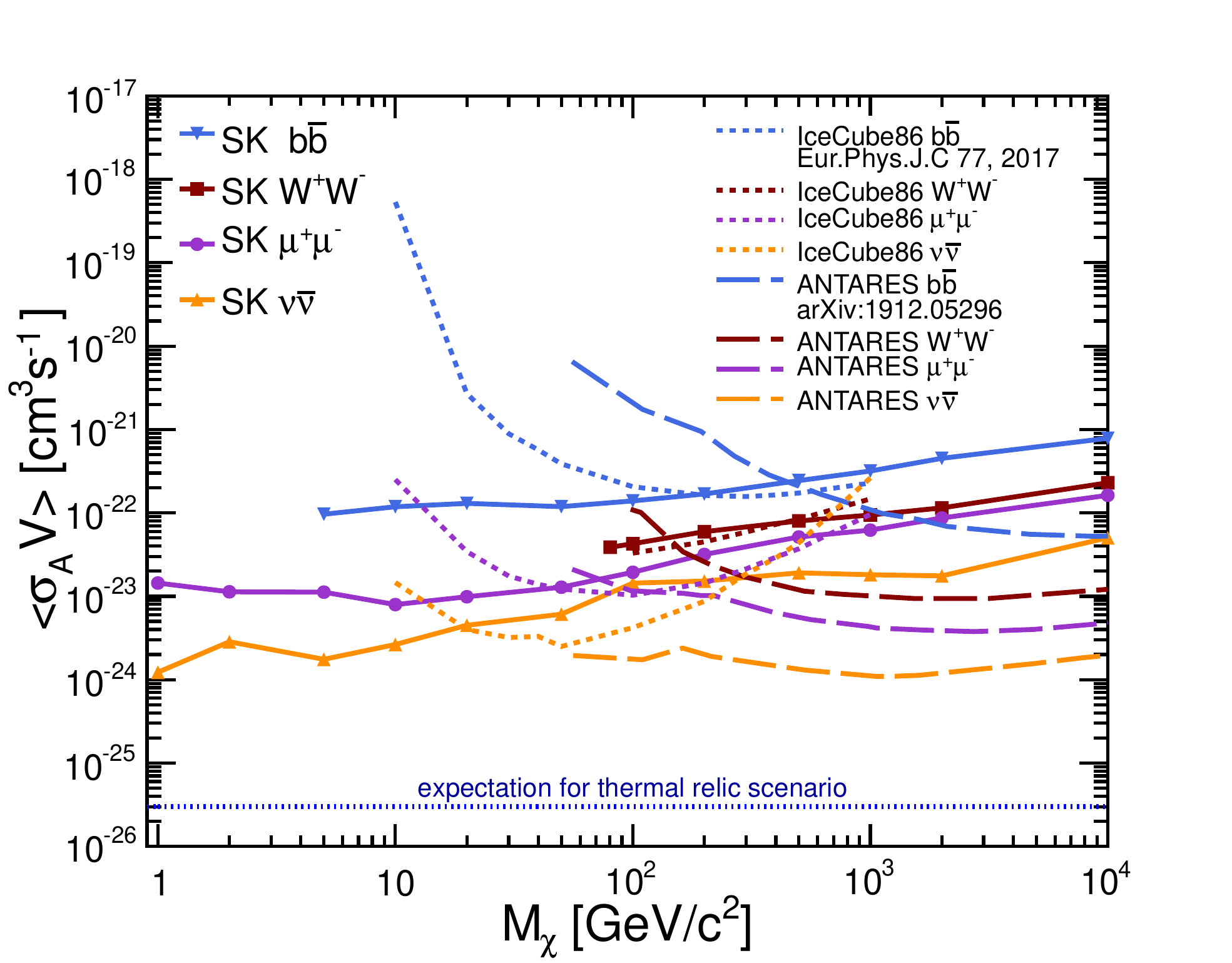}
	\end{minipage}
	\caption{Upper limits at $90\%$ C.L. on the DM self-annihilation cross section $\langle\sigma_{A} V\rangle$ (the region above the lines is excluded) obtained in the combined fit. SK limits for $\nu\bar \nu$,  $b\bar b$,  $W^{+}W^{-}$,  $\mu^{+}\mu^{-}$ obtained for the NFW halo profile are compared to results from IceCube~\cite{icecube2017} and ANTARES~\cite{antares2019}.}
	\label{fig:limit}
\end{figure*}

%%%%%%%%%%%%%%%%%%%%%%%%%%%%%%%%%%%%%%%%%%%%%%%%%%

An illustration of the samples used in the fit is shown in Fig.~\ref{fig:sk_ilustr}. 
The samples in the first column use only one angular bin and several momentum bins due to the poor angular resolution at low neutrino energy and lepton momentum. All other samples use multiple angular and momentum bins, though the latter have been merged together in the figure. 
The data are simultaneously fit in all neutrino flavors. 
Tau neutrinos end up mainly categorized as multi-GeV single-ring and multi-ring $e$-like events due to the complex final state of secondary particles produced in $\nu_{\tau}$ interactions and $\tau$ decay. 
Depending on $M_{\chi}$ and the annihilation mode, the signal appears in different samples of~Fig.~\ref{fig:sk_ilustr}. 
Even samples containing no signal contribution constrain the background rate and associated systematic errors, which impacts all bins in the analysis.

%%%%%%%%%%%%%%%%%%%%%%%%%%%%%%%%%%%%%%%%%%%%%%%%%%

%%%%%%%%%%%%%%%%%%%%%%%%%%%%%%%%%%%%%%%%%%%%%%%%%%
\section{Results of Combined Fit Analysis}
\label{sec:results}
%%%%%%%%%%%%%%%%%%%%%%%%%%%%%%%%%%%%%%%%%%%%%%%%%%

For each tested WIMP mass and annihilation channel, we try to determine the value of $\beta$ that governs the allowed contribution to the SK data from the simulated signal. The fitted number of DM-induced neutrino events accounting for the ``best-fit'' signal normalization $\beta$ is shown in Fig.~\ref{fig:allowed} along with the expected limits as calculated by MC under the assumption of no WIMP signal (denoted as $90\%$ and $99\%$ C.L. sensitivities). 

It should be noted that the points in Fig.~\ref{fig:allowed} are not independent, as the same set of data is used for each $M_{\chi}$ hypothesis. Accordingly, a correlated rise is seen in all annihilation channels at low WIMP masses as parts of the annihilation spectra from neighboring $M_{\chi}$ overlap in some analysis bins. At higher WIMP masses the results fall into the unphysical region, where the fitted number of WIMPs is negative.  A strong excess across all annihilation channels that exceeds the background-only expectation around a given WIMP mass is expected for a real signal. However, there is no such signal contribution allowed by the data. 

The final fit result is translated into an upper $90\%$ C.L. limit on the fitted number of DM-induced neutrinos using a Bayesian approach~\cite{pdg1996} 
to adjust results which fall into the unphysical region. Figure~\ref{fig:allowed_brazil} shows the result together with the sensitivity assuming no WIMP contribution. The $1-2\sigma$ uncertainty was derived from pseudo MC data sets constructed without any WIMP-induced neutrinos.

Prior to fitting the data, MC studies assuming a WIMP signal were used to determine a significance threshold for the analysis. 
An excess above the atmospheric background with a local $p$-value greater than $3\sigma$ was chosen as the criterion for a possible WIMP signal. 
Figure~\ref{fig:pvalue} shows the $p$-value distributions for the obtained fit results. There is a $\sim2\sigma$ excess observed in the $M_{\chi}=5$~GeV $b \bar b$ channel ($2.08\sigma$) , for the $M_{\chi}=1$~GeV $\mu^{+}\mu^{-}$ ($1.74\sigma$) channel, and for the $M_{\chi}=2$~GeV $\nu \bar \nu$ ($1.82\sigma$) channel. Though all $p$-values are consistent with no WIMP contribution, additional checks were performed for the most significant result.

A fit in which the position of the GC was treated as a free parameter moving in right ascension (RA), but fixed in declination (DEC), was performed for the $b\bar b$ annihilation channel and $M_{\chi}=5$~GeV. Fixing the declination allows us to use the same systematic uncertainties, as they depend on the zenith angle, which itself is proportional to the declination. This additional analysis found a similar signal excess of approximately $2\sigma$ for a WIMP signal, but for a GC position of RA in range from $210^\circ$ to $260^\circ$, though the true GC position is at $266^\circ$. Accordingly, we find no indication of a signal consistent with the expectation for DM from the GC halo in this channel.
   
Based on the limit on the number of DM-induced neutrinos, the corresponding limit on the diffuse flux 
is derived as a function of $M_{\chi}$ using Eq.~\ref{dmannihilationrate} and translated into a limit on the DM self-annihilation cross section $\langle\sigma_{A} V \rangle$. 
The latter is shown in~Fig.~\ref{fig:limit} and compared to limits obtained by other neutrino experiments. The line at the $\langle\sigma_{A} V\rangle = 3 \cdot 10^{-26}$~cm$^3$ s$^{-1}$ is the expectation for WIMPs produced thermally during the evolution of the Universe~\cite{bertone}. Despite the smaller effective area of the SK detector when compared to the IceCube detector~\cite{icecube2017}, the limits obtained in this analysis are stronger due to the fact that the SK detector can probe the GC with upward-going events. At the location of the SK detector, the GC is below the horizon for $\sim64\%$ of the year, while for IceCube it is always above the horizon and can only be directly probed with downward-going events which typically suffer from more backgrounds from cosmic ray muons. ANTARES is operating in the same hemisphere as SK and its limits~\cite{antares2019} are stronger than the ones obtained here for $M_{\chi}>50-500$~GeV (depending on the annihilation mode) due to the larger effective area of its detector. Weaker constraints from ANTARES observed for $M_{\chi}<500$~GeV for $b\bar b$ and for the $M_{\chi}<100-150$~GeV for $W^{+}W^{-}$/$\mu^{+}\mu^{-}$ annihilation channels are due to the different detection thresholds between ANTARES and SK. In this $M_{\chi}$ range, a substantial part of the DM-induced neutrino signal is expected below several tens of GeV (cf.~Fig~\ref{fig:dn_de}), a region that is well covered by the SK detector. 

%%%%%%%%%%%%%%%%%%%%%%%%%%%%%%%%%%%%%%%%%%%%%%%%%%
\section{On-source Off-source analysis}
\label{sec:on_off}
%%%%%%%%%%%%%%%%%%%%%%%%%%%%%%%%%%%%%%%%%%%%%%%%%%

\begin{figure}[t]
	\begin{minipage}{3.6in}
		\includegraphics[width=3.6in,keepaspectratio=true,type=pdf,ext=.pdf,read=.pdf]{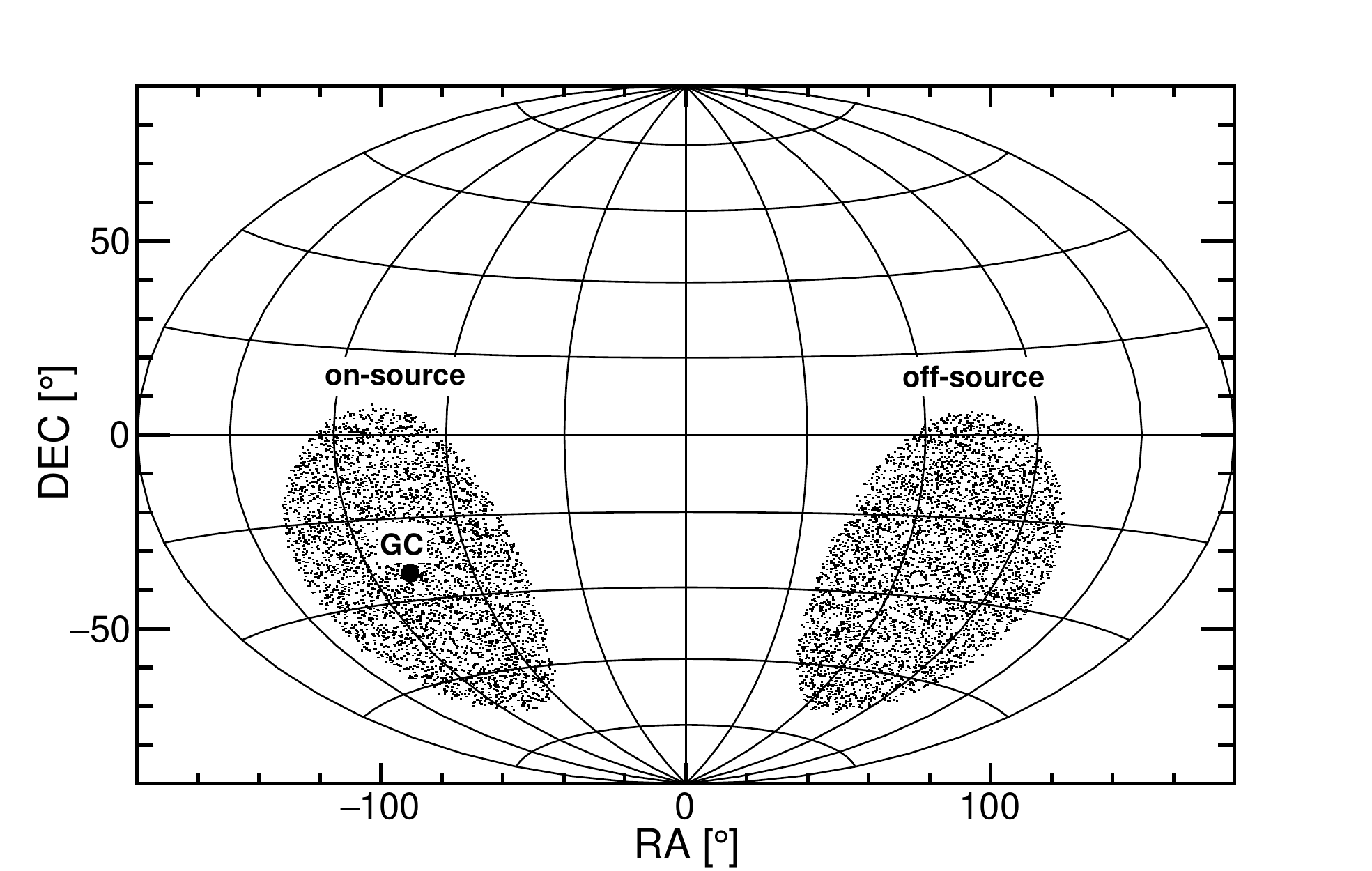}
	\end{minipage}
	\caption{The on-source off-source regions defined in equatorial coordinates.} 
	\label{fig:idea_on_off}
\end{figure}

The on-source off-source analysis provides a data-driven cross check of the analysis shown in the previous section, albeit with weaker sensitivity. 
It has the advantage of being able to estimate the background directly from data. 
Equally sized on- and off-source regions are defined in right ascension and declination as shown in Fig.~\ref{fig:idea_on_off}. 
Most of the signal is expected to come from the on-source region centered around the GC ($266^\circ$ RA,$-29^\circ$ DEC) 
while an independent background estimation is obtained from the off-source region, which is offset $180^\circ$ in right ascension but at the same declination. Note that the atmospheric neutrino background is expected to be the same in both regions as they 
correspond to identical zenith angles in SK's local coordinate system. 
Therefore, the expected number of events in the on-source region can be interpreted as $N_{ON} = N^{bkg}_{on} + N^{sig}_{on}$, 
while for the off-source it is $N_{OFF} = N^{bkg}_{off} + N^{sig}_{off}$.  
Here $N^{sig} (N^{bkg}$) stands for the number of signal (background) neutrinos in the on-source (off-source) regions. 
In this analysis we subtract the number of off-source events from the on-source observation: $N_{ON}-N_{OFF} = N^{sig}_{on}-N^{sig}_{off}$. 
This number effectively equals $N^{sig}_{on}$ as $N^{sig}_{off}$ is expected to be significantly smaller than $N^{sig}_{on}$ assuming a true source from the GC halo. The result of this subtraction is directly proportional to $\langle\sigma_{A} V\rangle$.
Systematic uncertainties related to the background should equally effect the on- and off-source regions and therefore cancel in the subtraction.

%%%%%%%%%%%%%%%%%%%%%%%%%%%%%%%%%%%%%%%%%%%%%%%%%%
%%%%%%%%%%%%%%%%%%%%%%%%%%%%%%%%%%%%%%%%%%%%%%%%%%

The angular size of the on- and off-source regions is determined to maximize $S/\sqrt{B}$, where $S$ stands for the number of expected signal events
and $B$ for the number of background events. This optimization is performed with the same signal and background Monte Carlo used in the combined fit. 
As the angular resolution of an event depends on the neutrino energy, the optimal size of the on- and off-source regions 
differs between the FC sub-GeV, FC multi-GeV, PC and UP-$\mu$ event samples as shown in Table~\ref{table:cones} for the three considered halo profiles.
%%%%%%%%%%%%%%%%%%%%%%%%%%%%%%%%%%%%%%%%%%%%%%%%%%
\begin{table}[ht]
	\begin{center}
		\small{ 
			\begin{tabular}{| l | l  l  l | } \hline
				Event class	& Optimal size $[^\circ]$ & & \\
				& NFW & Moore & Kravtsov \\
				\hline \hline
				FC sub-GeV	& 60 & 60 & 60 \\
				FC multi-GeV &  30 & 25 & 55 \\
				PC 	& 20 & 10 & 45 \\
				UP-$\mu$ & 10 & 5 & 40 \\
				\hline
				
			\end{tabular}
		}
	\end{center}
	\caption{Optimal size of on-source and off-source regions assuming DM annihilation for the NFW, Moore and Kravtsov halo profiles.} 
	\label{table:cones}
\end{table}
%%%%%%%%%%%%%%%%%%%%%%%%%%%%%%%%%%%%%%%%%%%%%%%%%%
%%%%%%%%%%%%%%%%%%%%%%%%%%%%%%%%%%%%%%%%%%%%%%%%%%
\begin{figure}[tb]
	\begin{minipage}{3.6in}
		\includegraphics[width=3.6in,type=pdf,ext=.pdf,read=.pdf]{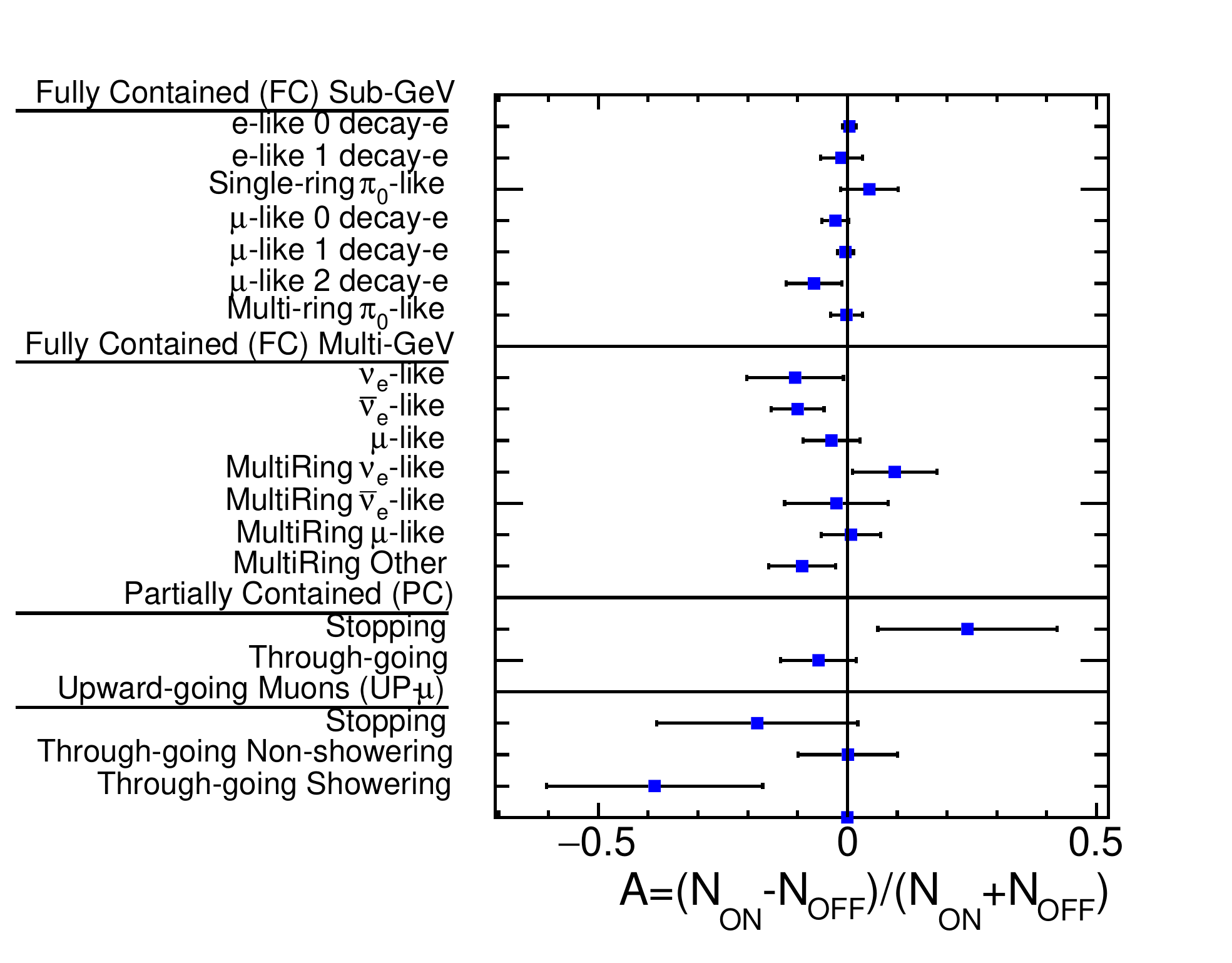}
	\end{minipage}
	\caption{Observed asymmetry between the number of neutrino events in the on- and off-source regions across the event subcategories for the NFW halo model. Errors on the points are statistical.}
	\label{fig:on_off}
\end{figure}
%%%%%%%%%%%%%%%%%%%%%%%%%%%%%%%%%%%%%%%%%%%%%%%%%%
%%%%%%%%%%%%%%%%%%%%%%%%%%%%%%%%%%%%%%%%%%%%%%%%%%
\begin{figure}[htb]
	\begin{minipage}{3.6in}
		\includegraphics[width=3.6in,type=pdf,ext=.pdf,read=.pdf]{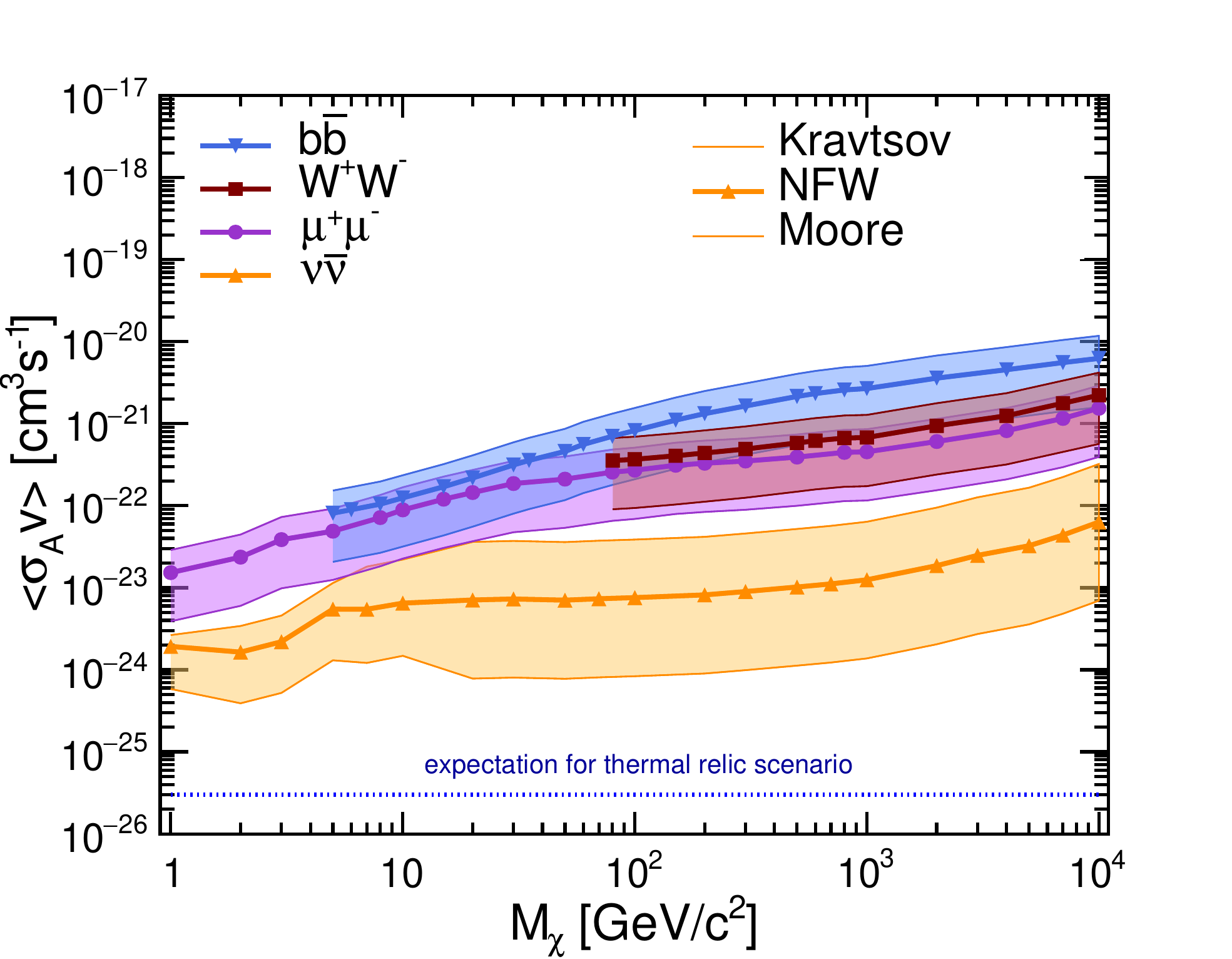}
	\end{minipage}
	\caption{Upper limits at $90\%$ C.L. on the DM self-annihilation cross section $\langle\sigma_{A} V\rangle$ as a function of
 $M_{\chi}$ for the $b\bar b$ (blue), $W^{+}W^{-}$ (maroon), $\mu^{+}\mu^{-}$ (purple) and $\nu\bar \nu$ (orange) annihilation channels as obtained in the on-source off-source analysis. The influence of the halo model choice is shown as bands around the results for the benchmark NFW profile.}
	\label{fig:on_off_limits}
\end{figure}
%%%%%%%%%%%%%%%%%%%%%%%%%%%%%%%%%%%%%%%%%%%%%%%%%%
%%%%%%%%%%%%%%%%%%%%%%%%%%%%%%%%%%%%%%%%%%%%%%%%%%
\begin{figure}[htb]
	\begin{minipage}{3.6in}
		\includegraphics[width=3.6in,type=pdf,ext=.pdf,read=.pdf]{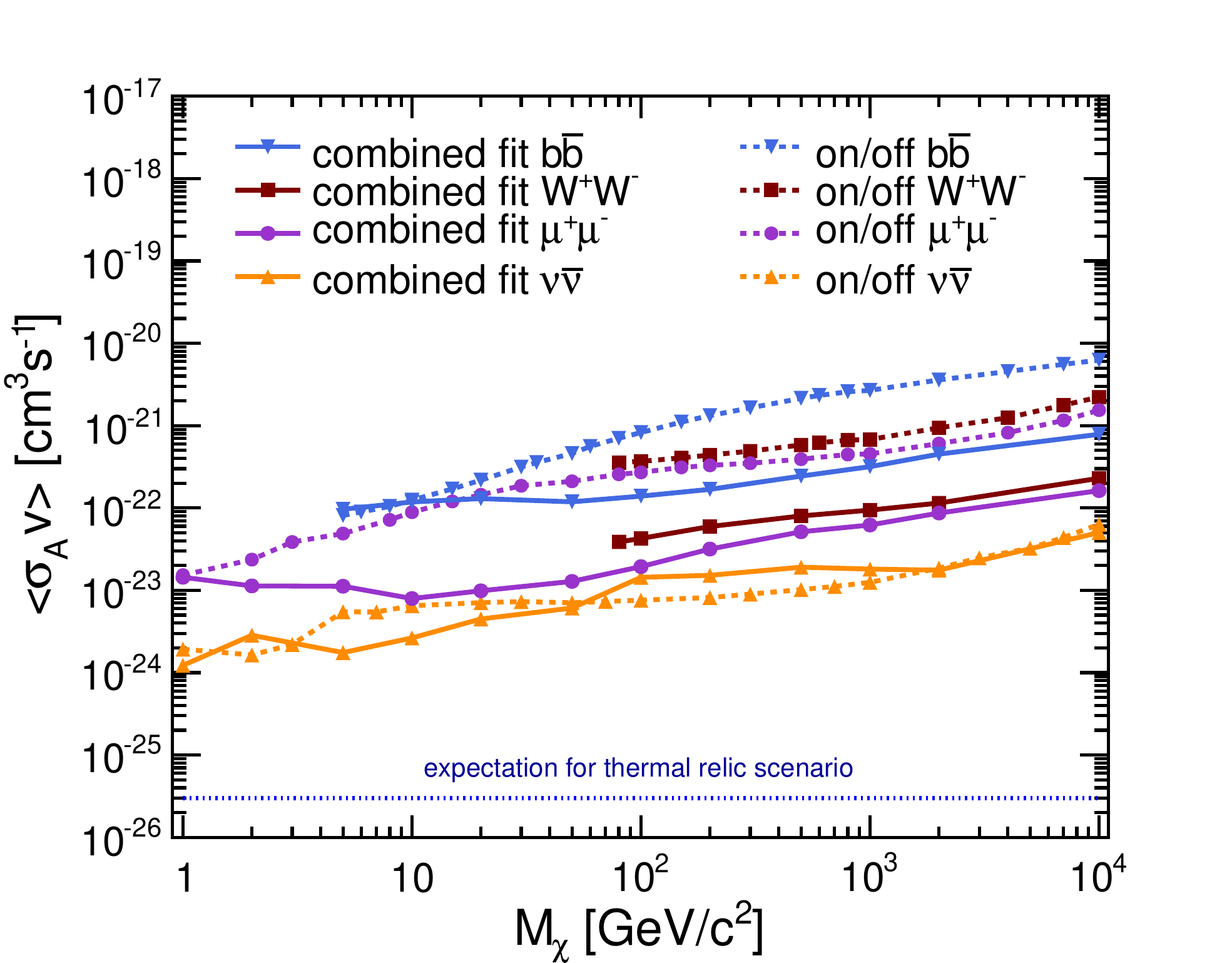}
		\end{minipage}
		\caption{Upper limits at $90\%$ C.L. on the DM self-annihilation cross section $\langle\sigma_{A} V\rangle$ as a function of $M_{\chi}$ for the combined fit (solid) and on-source off-source (dashed) analyses for the $b\bar b$ (blue), $W^{+}W^{-}$ (maroon), $\mu^{+}\mu^{-}$ (purple) and $\nu\bar \nu$ (orange) annihilation channels assuming NFW profile.}
		\label{fig:onoff_global}
	\end{figure}
%%%%%%%%%%%%%%%%%%%%%%%%%%%%%%%%%%%%%%%%%%%%%%%%%%
%%%%%%%%%%%%%%%%%%%%%%%%%%%%%%%%%%%%%%%%%%%%%%%%%%

The number $N_{ON}-N_{OFF}$ is obtained for each subcategory of events and 
similar numbers of events are observed in both the on- and off-source regions for all classes. 
Figure~\ref{fig:on_off} shows the asymmetry, $A = (N_{ON}-N_{OFF})/(N_{ON}+N_{OFF})$, for the benchmark NFW model and all event categories.

As no asymmetry in the event rate is observed, we constrain $\langle\sigma_{A} V\rangle$ 
by introducing information on the halo model, the mass of the DM particles and the annihilation channel.  
Limits are calculated for $M_{\chi}$ in the range from 1 GeV to 10 TeV for the $b\bar b$, $\mu^{+} \mu^{-}$, $W^{+}W^{-}$ and $\nu\bar \nu$ annihilation channels as shown in Fig.~\ref{fig:on_off_limits}. 
Differences between the NFW model and the Moore and Kravtsov models can reach an order of magnitude, 
due to large discrepancies in the predicted densities of the DM particles in the inner parts of the galaxy.

Figure~\ref{fig:onoff_global} compares the results of this analysis with the constraints from the combined fit presented in Section~\ref{sec:results}. 
The combined fit analysis yields limits that are roughly one order of magnitude stronger than the on-source off-source approach for $b\bar b$, $\mu^{+} \mu^{-}$ and $W^{+}W^{-}$. However, for low WIMP masses, the obtained limits are of similar strength, as the signal is mostly present in the sub-GeV samples. The pointing resolution of these samples is poorer than their higher energy counterparts which reduces their contribution to the sensitivity of the combined fit. 

For the $\nu\bar \nu$ mode the limits in both analyses are similar due to the fact that in the on-source off-source method 
the number $N_{ON}-N_{OFF}$ used in the limit calculation is based only on the sample with the expected DM contribution for the assumed 
mass. 
This effectively resembles the situation in the combined fit where the $\nu\bar \nu$ signal also appears in only a limited number of samples.
In contrast, the limit calculation for $b\bar b$, $\mu^{+} \mu^{-}$ and $W^{+}W^{-}$ is based on all samples together since the signal 
is broadly distributed for these channels.

%%%%%%%%%%%%%%%%%%%%%%%%%%%%%%%%%%%%%%%%%%%%%%%%%%
\section{Summary}
\label{sec:summary}
%%%%%%%%%%%%%%%%%%%%%%%%%%%%%%%%%%%%%%%%%%%%%%%%%%
The analyses presented above show no excess of DM-induced neutrinos from the galactic center or its halo; the data are consistent with the expectation from atmospheric neutrino backgrounds. 
The strongest exclusion is obtained for WIMP masses below several tens of GeV in the $b\bar b$, $\nu \bar \nu$, $\mu^{+} \mu^{-}$ channels in the combined fit. 
Limits on $\langle\sigma_{A} V\rangle$ reach as low as $9.6 \times10^{-23}$~cm$^3$ s$^{-1}$ for 5~GeV WIMPs in the $b\bar b$ channel and $1.2 \times10^{-24}$~cm$^3$ s$^{-1}$ for 1~GeV WIMPs in the $\nu \bar \nu$ channel.
These are the strongest indirect limits among similar WIMP-induced neutrino searches to date. 
Strong constraints on GeV-scale $M_{\chi}$ reflect the long exposure and the discrimination power of multiple subsamples reconstructed using the Super-Kamiokande detector. 
A complementary and completely data-driven result which compared event rates from the angular region near the GC and one of the same size but located in the opposite direction also found no evidence of an excess beyond the background expectation. 

%%%%%%%%%%%%%%%%%%%%%%%%%%%%%%%%%%%%%%%%%%%%%%%%%%
\section{Acknowledgments}
\label{sec:acknowledgments}
%%%%%%%%%%%%%%%%%%%%%%%%%%%%%%%%%%%%%%%%%%%%%%%%%%

We gratefully acknowledge the cooperation of the Kamioka Mining and Smelting Company. The Super-Kamiokande experiment has been built and operated from funding by the Japanese Ministry of Education, Culture, Sports, Science and Technology, the U.S. Department of Energy, and the U.S. National Science Foundation. Some of us have been supported by funds from the National Research Foundation of Korea NRF-2009-0083526 (KNRC) funded by the Ministry of Science, ICT, and Future Planning and the the Ministry of Education (2018R1D1A3B07050696, 2018R1D1A1B07049158), the Japan Society for the Promotion of Science, the National Natural Science Foundation of China under Grants No.~11235006, the Spanish Ministry of Science, Universities and Innovation (grant PGC2018-099388-B-I00), the Natural Sciences and Engineering Research Council (NSERC) of Canada, the Scinet and Westgrid consortia of Compute Canada, the National Science Centre, Poland (2015/18/E/ST2/00758), the Science and Technology Facilities Council (STFC) and GridPPP, UK, the European Union's H2020-MSCA-RISE-2018 JENNIFER2 grant agreement no.822070, and H2020-MSCA-RISE-2019 SK2HK grant agreement no. 872549.

%%%%%%%%%%%%%%%%%%%%%%%%%%%%%%%%%%%%%%%%%%%%%%%%%%
%%%%%%%%%%%%%%%%%%%%%%%%%%%%%%%%%%%%%%%%%%%%%%%%%%

%%%%%%%%%%%%%%%%%%%%%%%%%%%%%%%%%%%%%%%%%%%%%%%%%%

\end{document}